\documentclass[preprint]{elsarticle}
\usepackage{units}
\usepackage{upgreek}
\usepackage{amssymb}
\usepackage{textcomp}
\usepackage{gensymb}
\usepackage{lineno}

\journal{Nuclear Instruments and Methods in Physics Research Section A}

\begin{document}

\begin{frontmatter}

\title{A High-resolution Scintillating Fiber Tracker With Silicon
Photomultiplier Array Readout}
\author[AC]{B.~Beischer}
\author[AC]{H.~Gast\fnref{label1}}
\author[AC]{R.~Greim}
\author[AC]{W.~Karpinski}
\author[AC]{T.~Kirn}
\author[LA]{T.~Nakada}
\author[AC]{G.~Roper Yearwood}
\author[AC]{S.~Schael}
\author[AC]{M.~Wlochal}

\address[AC]{I. Physikalisches Institut B, RWTH Aachen University, 52074 Aachen, Germany}
\address[LA]{Ecole Polytechnique F\'ed\'erale de Lausanne, Dorigny, 1015 Lausanne, Switzerland}
\fntext[label1]{Present address: Max-Planck-Institut f\"ur Kernphysik, Saupfercheckweg 1,
69117 Heidelberg, Germany. Tel.: +49~6221~516-634, Fax: +49~6221~516-603, Email: henning.gast@mpi-hd.mpg.de}

\begin{abstract}
We present prototype modules for a tracking detector consisting of
multiple layers of $\unit[0.25]{mm}$ diameter scintillating fibers
that are read out by linear arrays of silicon photomultipliers. The module
production process is described and measurements of the key properties for both the fibers 
and the readout devices are shown.
Five modules have been subjected to a $\unit[12]{GeV}/c$ proton/pion testbeam
at CERN. A spatial resolution of $\unit[50]{\upmu{}m}$ and light
yields exceeding $20$ detected photons per minimum ionizing particle have
been achieved, at a tracking efficiency of more than $\unit[98.5]{\%}$. Possible
techniques for further improvement of the spatial resolution are discussed.
\end{abstract}


\end{frontmatter}


\section{Introduction}
Scintillating fiber trackers~\cite{ref:scifirev} have been realized for a number of
experiments such as D\O~\cite{dzero}, MICE~\cite{mice},
CHORUS~\cite{chorus} and K2K~\cite{k2k}. Of these, the D\O\ Central
Fiber Tracker has achieved the best resolution of about
$\unit[0.1]{mm}$ using double layers of $\unit[0.835]{mm}$ thin fibers
read out by Visible Light Photon Counters
(VLPCs)\cite{vlpc}. VLPCs offer single-photon resolution due to a high
internal gain of about $10^4$ and are linear up to hundreds of
photons. The main drawback of VLPCs is that they have to be operated
at cryogenic temperatures which introduces a significant overhead into
the operation of scintillating fiber trackers as used in the D\O\ and
MICE experiments. The CHORUS fiber tracker achieves a spatial
resolution of $\unit[0.185]{mm}$ for a single ribbon of 7 layers of
$\unit[0.5]{mm}$ thin scintillating fibers and uses CCD cameras instead of
VLPCs for readout, as does the K2K scintillating fiber tracker. CCD
cameras require image intensifiers to detect the few tens of photons
that thin scintillating fibers emit for a minimum ionizing
particle. Image intensifiers and photomultipliers are sensitive to
magnetic fields and require operating voltages of a few thousand
volts.

A new type of scintillating fiber tracker has become possible with the
advent of silicon photomultipliers~\cite{ref:sipm1,ref:sipm2,ref:sipm_buzhan,ref:sipm3,ref:sipm_renker,ref:sipm_danilov} (SiPMs) as a viable
alternative for the commonly used photon detectors. SiPMs like VLPCs
have a high intrinsic gain of $10^5 - 10^6$ but can be operated at
room temperature. Furthermore, they are insensitive to magnetic fields
and are operated at voltages of $\unit[20]{V} - \unit[80]{V}$. Another
key feature of SiPMs are their compact dimensions. This allows the
design of detector modules that have almost no dead area if integrated
SiPMs are used as photon detectors.

In this article, we describe the design of a new tracking device
aiming at the detection of charged particles with an efficiency above
$98.5\,\%$ and a spatial resolution of $\unit[0.05]{mm}$. It
consists of modules made of ribbons of $\unit[0.25]{mm}$ diameter 
scintillating fibers  that are read out by silicon
photomultiplier arrays. After an overview of the tracker and module
design, the process of module production as well as the optical hybrid
used for readout are described in section~\ref{sec:tracker}. Sections~\ref{sec:fibers}
and~\ref{sec:arrays} deal with the scintillating fibers and silicon
photomultiplier arrays employed, respectively. Finally, the performance 
of several prototype modules as determined in a testbeam at CERN is evaluated in
section~\ref{sec:testbeam}. Possible future improvements are discussed
in section~\ref{sec:outlook}.

\section{Tracker layout}
\label{sec:tracker}
\subsection{Tracker and module design}
\begin{figure}
 \centering
 \includegraphics[width=0.8\columnwidth]{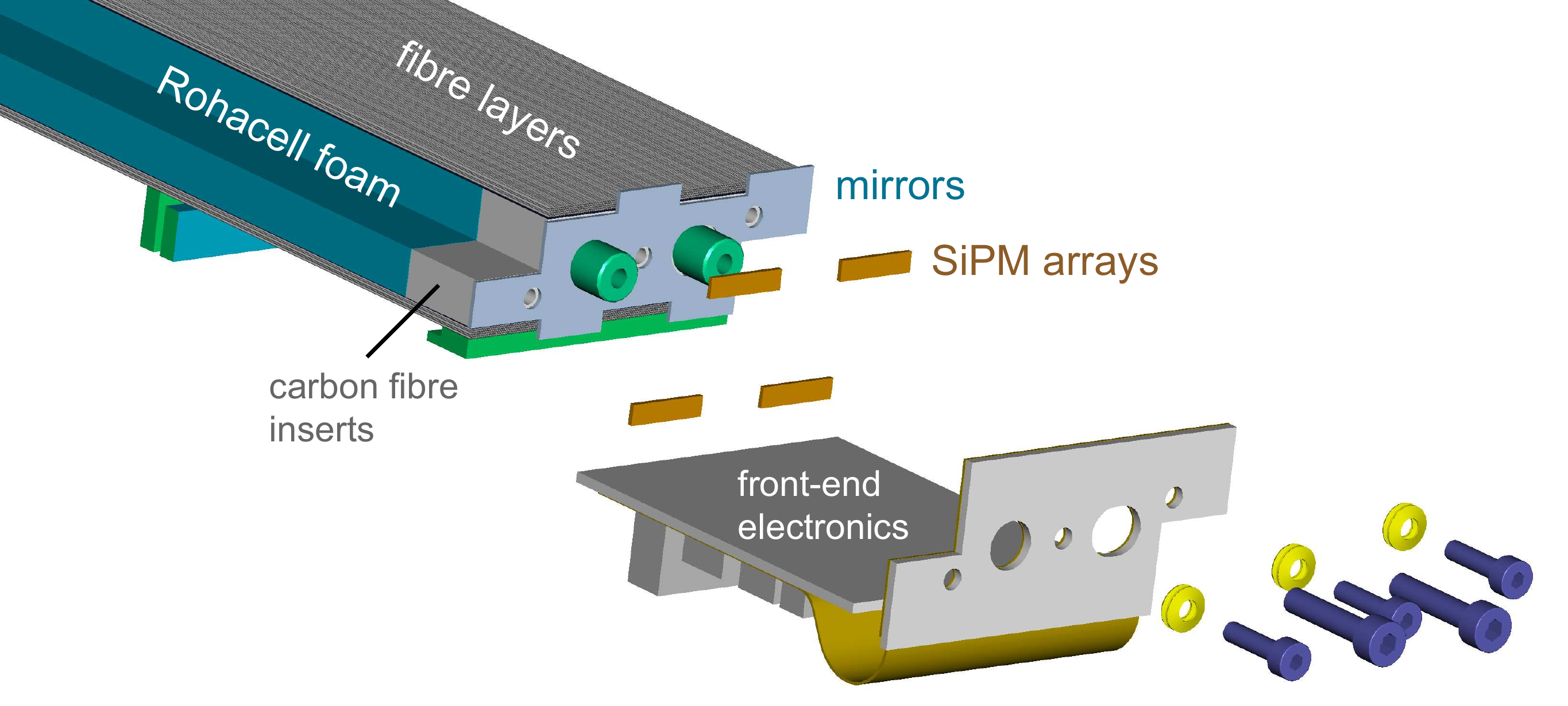}
 \caption{Exploded view of a tracker module. Two ribbons of
scintillating fibers located at the top and bottom are
carried by Rohacell foam with carbon fiber inserts at the
ends. Optical hybrids equipped with
silicon photomultiplier arrays and mirrors are screwed to the sides of the module. 
A corresponding hybrid is mounted to the far side of the module so that each fiber is covered
by an SiPM channel on one side and a mirror on the other. The large screws
will be used to mount the module to the tracker walls.}
 \label{fig:moduleschematic}
\end{figure}

The tracker described here is being developed for use in the PEBS~\cite{pebs}
balloon-borne detector. It consists of layers made up from modules like the one depicted in
figure~\ref{fig:moduleschematic}. Charged particles
traversing the module deposit energy in two ribbons of scintillating
fibers that are located at the top and bottom of the module, creating scintillation photons.
A small fraction of the scintillation light is
then guided by total internal reflection to the fiber ends where it is
detected by silicon photomultiplier arrays. Each module thus allows two
independent measurements of the intersection point of a
trajectory. The modules are Z-shaped which allows for them to be
placed closely next to each other so that there are no gaps in a
tracker layer. The overlap between the modules allows for internal 
alignment with tracks within one layer.
Each module consists of a
mechanical support made up from low-density ($50\,\mathrm{kg}/\mathrm{m}^3$) Rohacell foam between
two thin ($\unit[0.1]{mm}$) carbon fiber skins.
It carries scintillating fiber ribbons that are
$\unit[32]{mm}$ wide and up to $\unit[2]{m}$ long on both sides.
A fiber ribbon consists of five layers of
$\unit[0.25]{mm}$ thin scintillating fibers, glued together in the
tightest arrangement. Each module end holds two
precision pins for a controlled mounting of optical hybrids carrying
four SiPM arrays each.
The SiPM arrays have a $\unit[8]{mm}$ wide and $\unit[1]{mm}$ high active area and are segmented
into 32 individual SiPMs with 80 pixels each, with a readout pitch of
$\unit[0.25]{mm}$, matching the
diameter of the fibers. Groups of 32 fiber columns are read out by an
SiPM array at alternating ends of the module. Due to the mounting 
of the SiPM arrays, sensors cannot be placed next to each other
without a dead area of $\unit[0.25]{mm}$ between them. To minimize this
dead zone, a mirror covers the space between
the SiPMs in order to increase the light collection on the opposite
fiber end.

\subsection{Module production}
\label{sec:moduleprod}
Kuraray SCSF-81MJ fibers~\cite{kuraray} (sec.~\ref{sec:fibers}) with a
diameter of $\unit[0.25]{mm}$ are
used for the tracker modules. Ribbons of five
layers of 128 fibers each are produced in a
winding process similar to the one used to produce the fiber trackers
for the CHORUS and K2K experiments. The scintillating fiber arrives
uncut on a spool from the manufacturer. A helical groove with a pitch of
$\unit[0.275]{mm}$, to accommodate small variations in the fiber thickness,
is cut into an aluminum drum of $\unit[200]{mm}$
diameter on a
winding machine. The diameter is a free parameter and
can be adapted to the required module length. 
A release agent is applied to the drum prior to
the ribbon production. One layer of scintillating fiber is then wound
on top of the drum, with its position precisely fixed by the helical
groove, using a controlled tension of $\unit[20]{g}$ and
EpoTek 301M is applied as an adhesive. The fiber end is
fixed with a fast curing adhesive and cut. The next fiber layer is
wound on top of the first layer so that the fiber lies in the gap
between fibers of the previous layer maintaining a constant fiber
pitch over the layers. This process is repeated applying an adhesive
to the fibers after each completed layer until five layers of fibers
have been wound on top of the drum. After the last fiber layer 
has been completed, two aluminum end-pieces are screwed next to each other 
onto the drum to ease the handling of the fiber ribbon. 
The drum continues to rotate at a constant speed for a few hours
until the adhesive is cured. The ambient temperature for the whole 
process is kept at $\unit[22]{^\circ{}\mathrm{C}}$. The fiber ribbon is then 
cut between the end-pieces, taken off the drum and placed in a Teflon
bed that is filled up with glue and then covered by a glass plate. 
After the glue has fully cured, the setup with the fiber ribbon is 
placed in an oven heated to approximately $\unit[50]{^\circ{}C}$ to 
allow the fiber ribbon to straighten out.

\begin{figure}
 \centering
 \includegraphics[width=0.8\columnwidth]{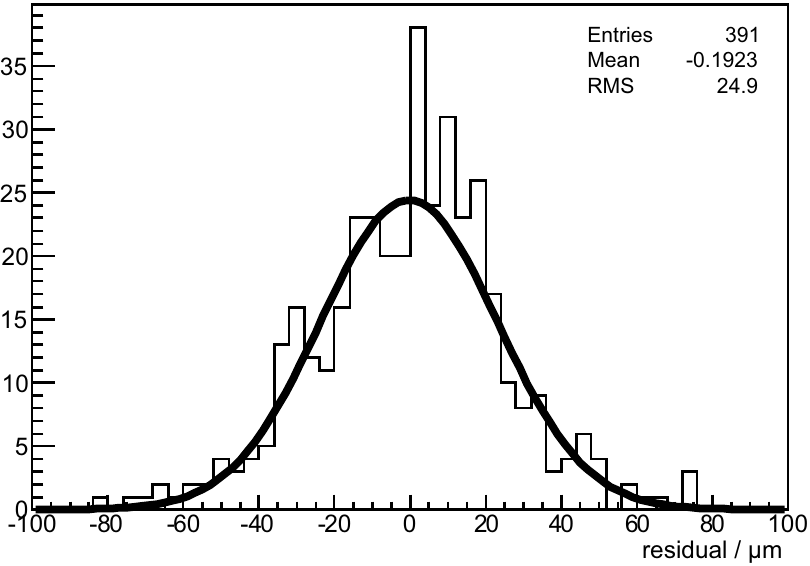}
 \caption{Distribution of the deviation of the fiber centers from a
uniform lattice with a pitch of $\unit[0.275]{mm}$, shown here for the
fibers included in the picture of fig.~\ref{fig:moduleEnd}. The black
line is a Gaussian fit to the data, plotted here for illustrative purposes.}
\label{fig:fiberlattice}
\end{figure}

\begin{figure}[htb]
\begin{center}
\includegraphics[width=\columnwidth,angle=0]{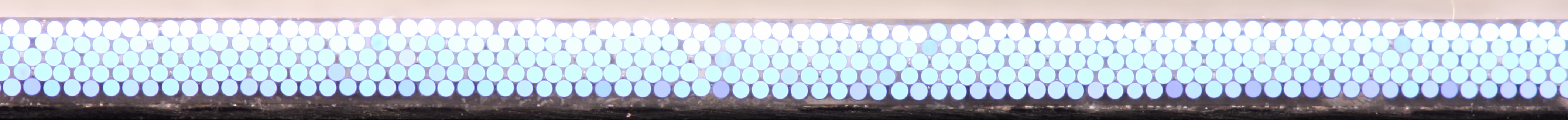}
\end{center}
\caption{Close-up photograph of a completed fiber ribbon made of Kuraray SCSF-81MJ fibers of $\unit[0.25]{mm}$ diameter.
The nominal gap in the horizontal direction is $\unit[0.025]{mm}$. Five layers of fibers
are placed in the tightest arrangement. The supporting carbon fiber skin is seen at the bottom of the picture.}
\label{fig:moduleEnd}
\end{figure}

The precision of the fiber placement in the ribbons is determined by scanning the
cross-section in the end-pieces to be better than $\unit[0.025]{mm}$ for a ribbon
consisting of 5 layers of about 128 fibers each
(fig.~\ref{fig:fiberlattice}). A photograph of the front of
a completed ribbon, illuminated from above, is shown in
figure~\ref{fig:moduleEnd}.

Two fiber ribbons are glued to the top and bottom of a mechanical
support made from a $\unit[10]{mm}$ thick Rohacell foam layer
contained between two $\unit[0.1]{mm}$ thin carbon fiber skins. 
Polycarbonate end-pieces are embedded in the support structure
to allow the mounting of the optical hybrid.
The module ends
are then polished to achieve a good optical coupling between the fibers and the SiPMs.
Finally, optical readout hybrids carrying the SiPMs (sec.~\ref{sec:hybrids}) on
a PCB board including the mirrors are screwed directly to the polished fiber ends on
both sides of the module.

\subsection{Optical hybrid and data acquisition}
\label{sec:hybrids}
\begin{figure}
\begin{center}
\includegraphics[width=0.8\columnwidth]{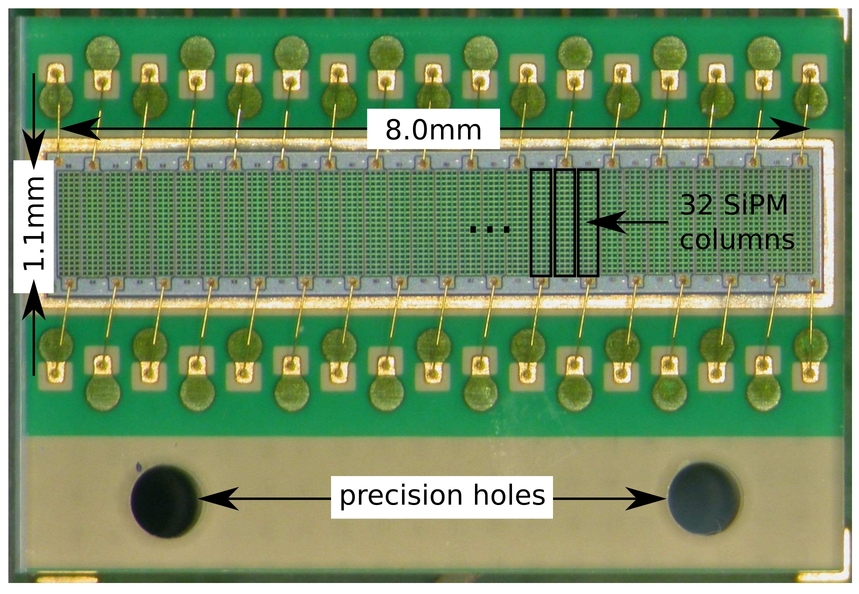}
\caption{Microscope picture of a Hamamatsu MPPC 5883 SiPM array
containing 32 individual SiPMs with a pitch of $\unit[0.25]{mm}$. The active area of each SiPM
measures $\unit[1.1]{mm}$ in height. The individual SiPMs are bonded towards alternating sides. Each SiPM
contains 80 pixels that are organized in 20~rows and four columns.}
\label{fig:mppc}
\end{center}
\end{figure}

The fiber modules are read out by SiPM arrays of type Hamamatsu MPPC 5883 (fig.~\ref{fig:mppc}). Each array 
consists of 32 independent SiPMs (called channels in the following)
and is mounted to a ceramic and protected by an epoxy layer above
the sensitive area of the semiconductor. The detector signals are read out
by bonding wires that are fed 
through to the backside of the ceramic. Section~\ref{sec:arrays} deals with a
detailed performance study of the SiPM arrays.

\begin{figure}
\begin{center}
\includegraphics[width=0.7\columnwidth]{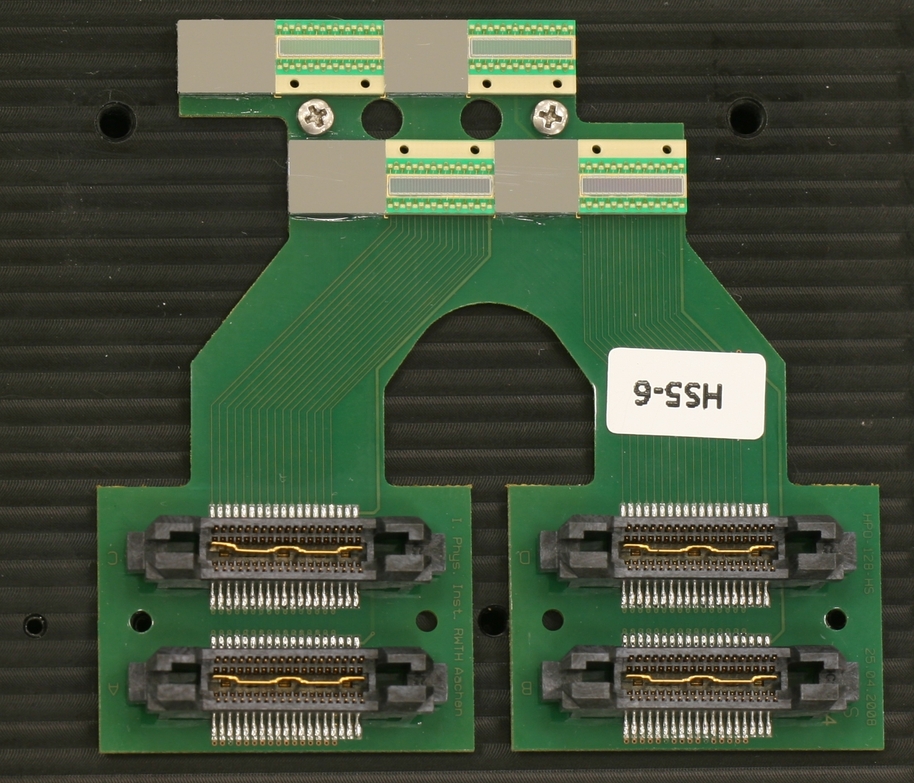}
\caption{Optical hybrid. Four SiPM arrays with mirrors in between are visible at the top of the picture,
together with the connectors to the front-end electronics in the lower part.}
\label{fig:opticalHybrid}
\end{center}
\end{figure}

Four arrays are positioned on an optical hybrid
(fig.~\ref{fig:opticalHybrid}) in the following way: The detectors are glued 
to a printed circuit board in a mechanical device for placing the detectors with an accuracy of $\unit[10]{\upmu{}m}$. 
The device is a flat surface equipped with four groups of two
neighboring positioning pins for placing the SiPM 
arrays in the geometry defined by the tracker module. The SiPM arrays are pinned up in their precision 
holes face down. A shadow mask with holes at the position of the contacts is put inside the device to 
brush a conductive glue of type Elecolit 323 into the holes and thus
on the contacts. After the mask has been removed, the PCB is pressed 
on the contacts with a defined pressure. After this procedure, mirrors are placed inside the gap between the
SiPM arrays to increase the light yield at the opposite side of a
fiber module. Two completed hybrids are screwed to the sides of a
fiber module.
The hybrid is read out by a specially developed front-end electronics board based on VA\_32/75 chips 
produced by IDEAS, Norway~\cite{ideas}. VA\_32/75 chips are charge sensitive preamplifiers with a sample and hold 
stage. They have 32 channels each with a shaping time of $\unit[75]{ns}$ and a linear range of 
$\unit[36]{fC}$~\cite{ideas}. Due to the gain of $\sim 10^6$ of SiPMs the signals have to be attenuated
in a resistor network on the front-end board by a factor of 150 to match the dynamic range of the chip.

An external trigger generates a hold signal for the VA chips on the
front-end board with a delay of the shaping time of the preamplifier, so
that the SiPM signals are sampled and their amplitudes stored on the VA\_32/75 chips for
readout. The stored analog signal heights are digitized sequentially
by 12 bit \unit[5]{MHz} sampling ADCs on back-end boards. The data are
subsequently transferred to a PC via a QuickUSB interface produced by
Bitwise Systems~\cite{quickUsb}.

\section{Scintillating fibers}
\label{sec:fibers}
Kuraray SCSF-81MJ fibers were chosen for the tracker because their
peak emission wavelength at $\unit[437]{nm}$ matches the peak
sensitivity of the employed SiPMs. The fibers are multiclad, consisting
of a scintillating fiber core with a refractive index of $1.59$
surrounded by two claddings with refractive indexes of $1.49$ and
$1.42$, respectively, which are approximately $\unit[10]{\upmu{}m}$
thin. Measurements of the key properties of the fibers are shown in
the following: A long attenuation length and uniform thickness are vital for use
in a large-area tracking device, while the angular emission spectrum
is needed for an understanding of the observed spatial resolution.

\subsection{Attenuation length}
\begin{figure}
 \centering
 \includegraphics[width=0.8\columnwidth]{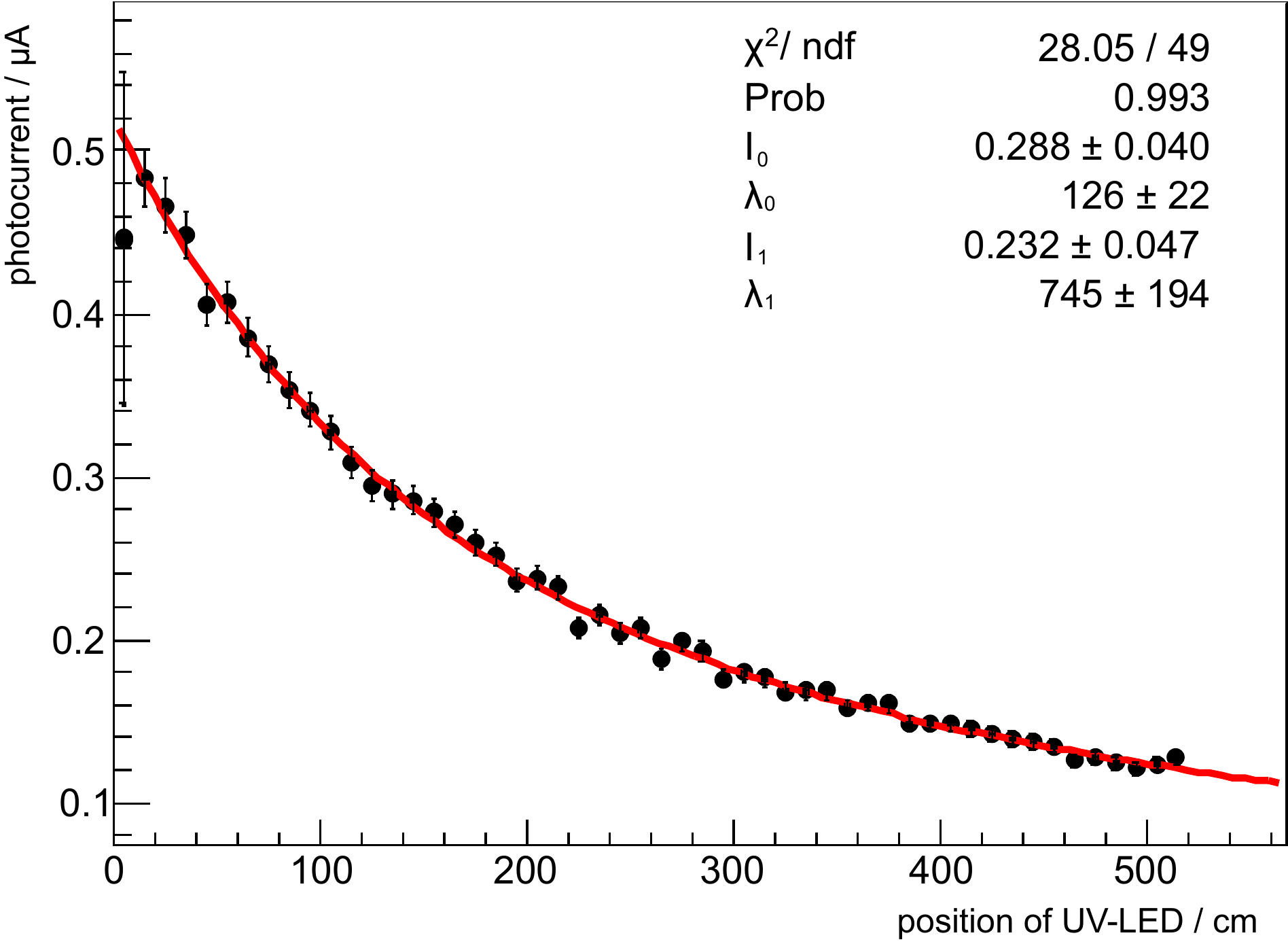}
 \caption{Determination of the absorption length of the scintillating fibers.
 The mean photo-current measured for four approximately $\unit[5]{m}$
long samples of fiber is shown. A UV-LED has been used to excite the fibers. The curve
is a fit according
to eq.~(\ref{eq:abslength}), and the fit parameters are given in the inset.}
 \label{fig:abslength}
\end{figure}

The manufacturer specifies an attenuation length of at least
$\unit[3.5]{m}$. We conducted a separate measurement of the
attenuation length for four $\unit[5]{m}$ long samples of Kuraray
fibers.
A UV-LED was used to excite the 
fiber locally while a photo-diode was used to measure the photo-current $I$ as a function of
position $x$ of the UV-LED
(fig.~\ref{fig:abslength}). A two-component exponential function
\begin{equation}
\label{eq:abslength}
I(x)= I_0\exp(-x/\lambda_\mathrm{0}) + I_1\exp(-x/\lambda_\mathrm{1})
\end{equation}
was fitted to the measured photo-current. The result shows that the light 
trapped within the fiber is made up of two components of roughly equal 
intensity. The first component has a short attenuation length of roughly
$\lambda_\mathrm{0}=\unit[(1.3\pm0.2)]{m}$ while the second component shows a longer attenuation 
length of $\lambda_\mathrm{1}=\unit[(7.5\pm2)]{m}$. The mean attenuation length over the first 
two meters of fiber, obtained from a fit to a single exponential, is 
approximately $\unit[(2.6\pm0.1)]{m}$. This is the figure that limits the module length.

\subsection{Thickness measurements}
\begin{figure}
 \centering
 \includegraphics[width=0.8\columnwidth]{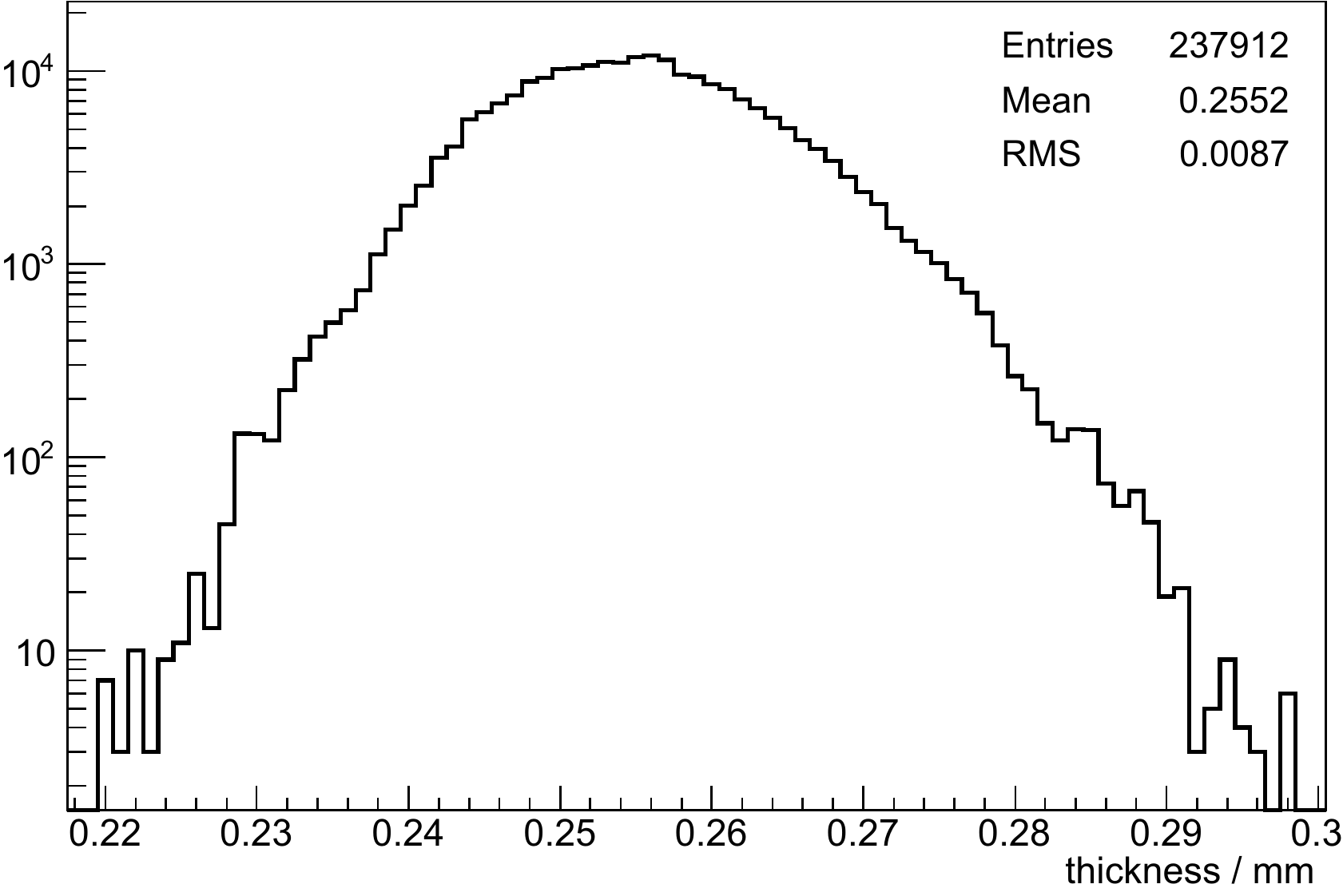}
 \caption{Quality assurance of the scintillating fibers. The thickness of the Kuraray SCSF-81MJ fibers
used in the module production process have continuously been sampled over \unit[3]{km} of fiber.
The nominal pitch of $\unit[0.275]{mm}$ accommodates the small variations in fiber thickness.}
 \label{fig:fiberdiameter}
\end{figure}

The thickness of the fiber was controlled by the manufacturer using a
micrometer on two $\unit[15]{m}$ long samples finding a mean thickness of
$\unit[0.252]{mm}$ and a root mean square variation of
$\unit[0.009]{mm}$.
The fiber is also measured prior to assembly
by a calibrated Zumbach ODAC-XY~\cite{zumbach} laser micrometer. The result of the measurement is shown in
figure~\ref{fig:fiberdiameter}. The fiber diameter of $(0.255\,\pm\,0.009)\,\mathrm{mm}$ is in good agreement with the measurement from
Kuraray. Furthermore, the simultaneous measurement in two directions
allows to study the ellipticity of the fiber and no deviation from a
circular shape was detected.

\subsection{Angular emission spectra}
\begin{figure}
 \centering
 \includegraphics[width=0.8\columnwidth]{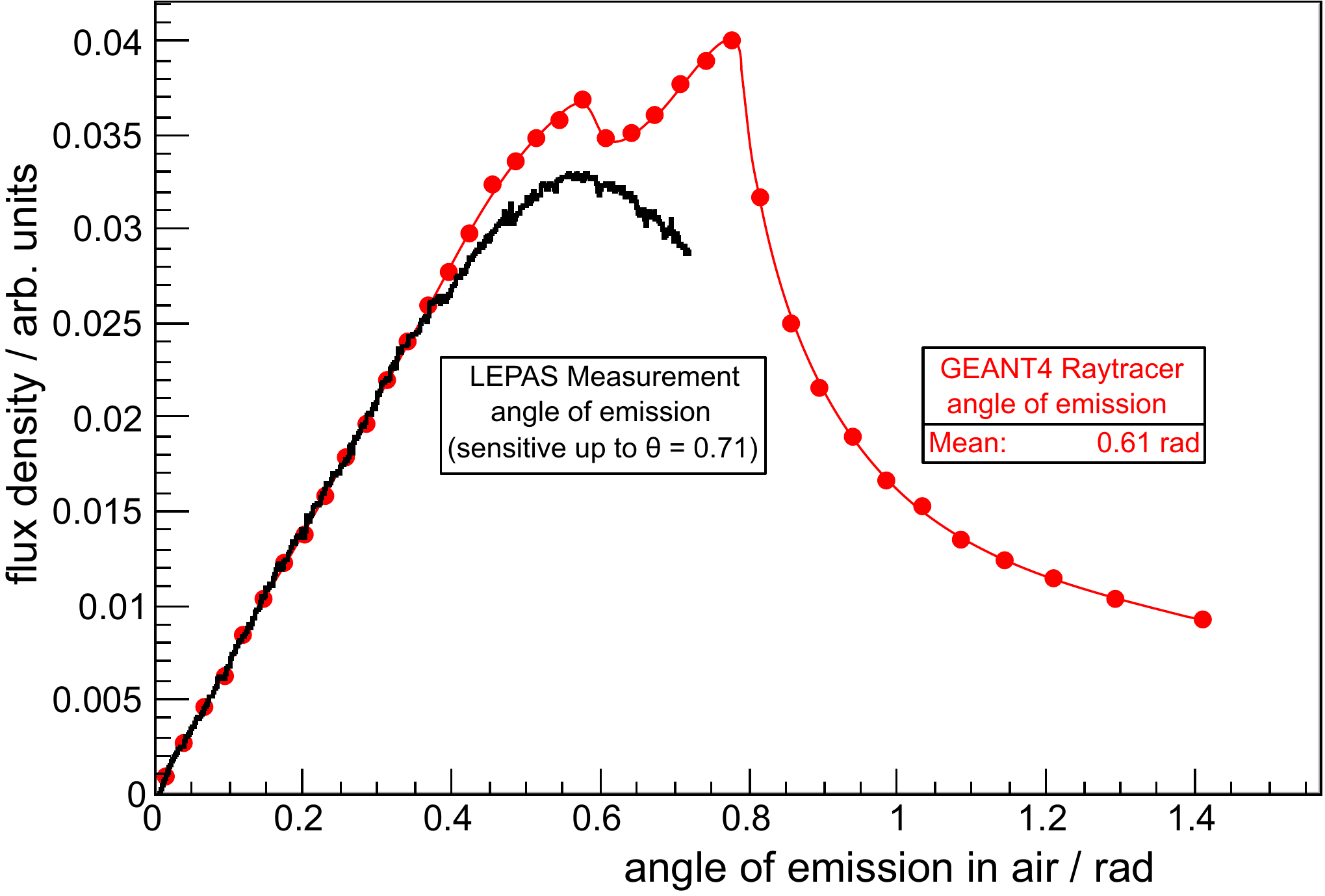}
 \caption{The angular distribution of photons emitted from the fiber
shows that the photons exit the fiber under an average angle of
$\unit[0.61]{rad}$ ($\unit[35]{^\circ{}}$) with respect to the fiber axis.}
 \label{fig:farfield}
\end{figure}

The angular emission spectrum of the fiber has been measured using a
Hamamatsu LEPAS optical beam measurement system~\cite{hamamatsu}. The setup records an image of the light emitted
from the fiber end and uses it to calculate the far-field. As its
angular range is limited to $\pm41^\circ$,
part of the cladding modes, which are trapped in the fiber by the outer
cladding, and helix modes, which are emitted from the fiber end under a
large angle of emission, could not be measured. The result is shown in
figure~\ref{fig:farfield}. It is compared to the prediction of a Monte
Carlo simulation based on Geant4~\cite{ref:g4} that traces the
trajectories of optical photons, based on an idealized cylindrical fiber geometry,
refractive indexes and absorption using geometric optics.
As shown in the figure, the simulation is in
agreement with the far-field measurement over the most part of the accessible angular range
and predicts an average angle of
emission of about $\unit[35]{^\circ{}}$ with respect to the fiber axis
in a medium matching the refractive index of the fiber core. The features 
at $\unit[35]{^\circ{}}$ and $\unit[46]{^\circ{}}$ in the Geant4 simulation
show the transitions from core modes trapped within the fiber core to
cladding modes that are totally reflected by the outer cladding and from cladding modes 
to helix modes that are trapped in the fiber region very close to the fiber claddings, respectively. 
The implication for the tracker module design is
that the silicon photomultipliers have to be mounted as close
to the fiber ends as technically possible. Any gap, for example from an protective epoxy layer
on top of the SiPM, significantly smears out the position information carried by the
photons. In this case photons emitted from a fiber will
not necessarily be detected by the SiPM array channels directly in
front of that fiber. As a consequence, the spatial resolution
deteriorates.

\section{Silicon photomultiplier arrays}
\label{sec:arrays}
A silicon photomultiplier (SiPM)~\cite{ref:sipm1,ref:sipm2,ref:sipm_buzhan,ref:sipm3,ref:sipm_renker,ref:sipm_danilov} is a
novel solid state photo-detector. In principle, it consists of a matrix
of avalanche photo diodes (APDs) that are operated in Geiger
mode, i.e.~above the breakdown voltage, and can be triggered by an
incident photon. The
single matrix elements are called pixels and the output signal is then
proportional to the number of pixels that have fired. This will in
turn be proportional to the number of incident photons as long as this
number is small compared to the number of pixels. The mode of
operation leads to a high intrinsic gain of the SiPM on the order of
$10^5$ to $10^6$. They have the virtues of high quantum efficiency, as well as
compactness, auto-calibration, and insensitivity to magnetic fields.
This allows for them to be used inside a particle spectrometer. 
The Hamamatsu MPPC 5883 SiPM arrays used for the tracker consist of 32
independent SiPMs ("channels")
with a pitch of $\unit[0.25]{mm}$. Each channel 
has $4\times{}20$ pixels providing enough dynamic range for minimum
ionizing particles (MIPs) of charge $z\lesssim{}4$ in the tracker, given that
a singly-charged MIP will trigger typically 10 pixels per channel
(sec.~\ref{sec:testbeam_results}). At high light yields,
saturation effects due to the limited number of pixels will lead to a
departure from the linearity between the number of triggered pixels
and the number of incident photons.

The key figure of merit of an SiPM is its photon detection efficiency $\epsilon_\mathrm{PDE}$
which can be expressed as the product

\begin{equation}
\label{eq:epde}
\epsilon_\mathrm{PDE} = \epsilon_\mathrm{ABE} \cdot
\epsilon_\mathrm{QE} \cdot \epsilon_\mathrm{FF}
\end{equation}

where $\epsilon_\mathrm{ABE}$ denotes the avalanche breakdown
efficiency, $\epsilon_\mathrm{QE}$ the
quantum efficiency of the pixels and $\epsilon_\mathrm{FF}$ the
fill factor that depends on the geometry of the SiPM and is the
critical factor in the determination of
$\epsilon_\mathrm{PDE}$. In the following, the method used for the
SiPM calibration is briefly reviewed in
section~\ref{sec:led}. Measurements of the fill factor and the photon
detection efficiency are
described in sections \ref{sec:ff} and \ref{sec:pde},
respectively. For a reliable operation of the tracker, the homogeneity
of the response of individual array channels is crucial and this is
explored in sec.~\ref{sec:homo}. Photons produced in the
avalanche process can cause neighboring pixels to fire. This crosstalk
effect is a potential obstacle in the operation of SiPMs. The
determination of the crosstalk probability is therefore the topic of sec.~\ref{sec:xtalk}.

\subsection{Analysis of LED spectra}
\label{sec:led}
\begin{figure}
\begin{center}
\includegraphics[width=\columnwidth]{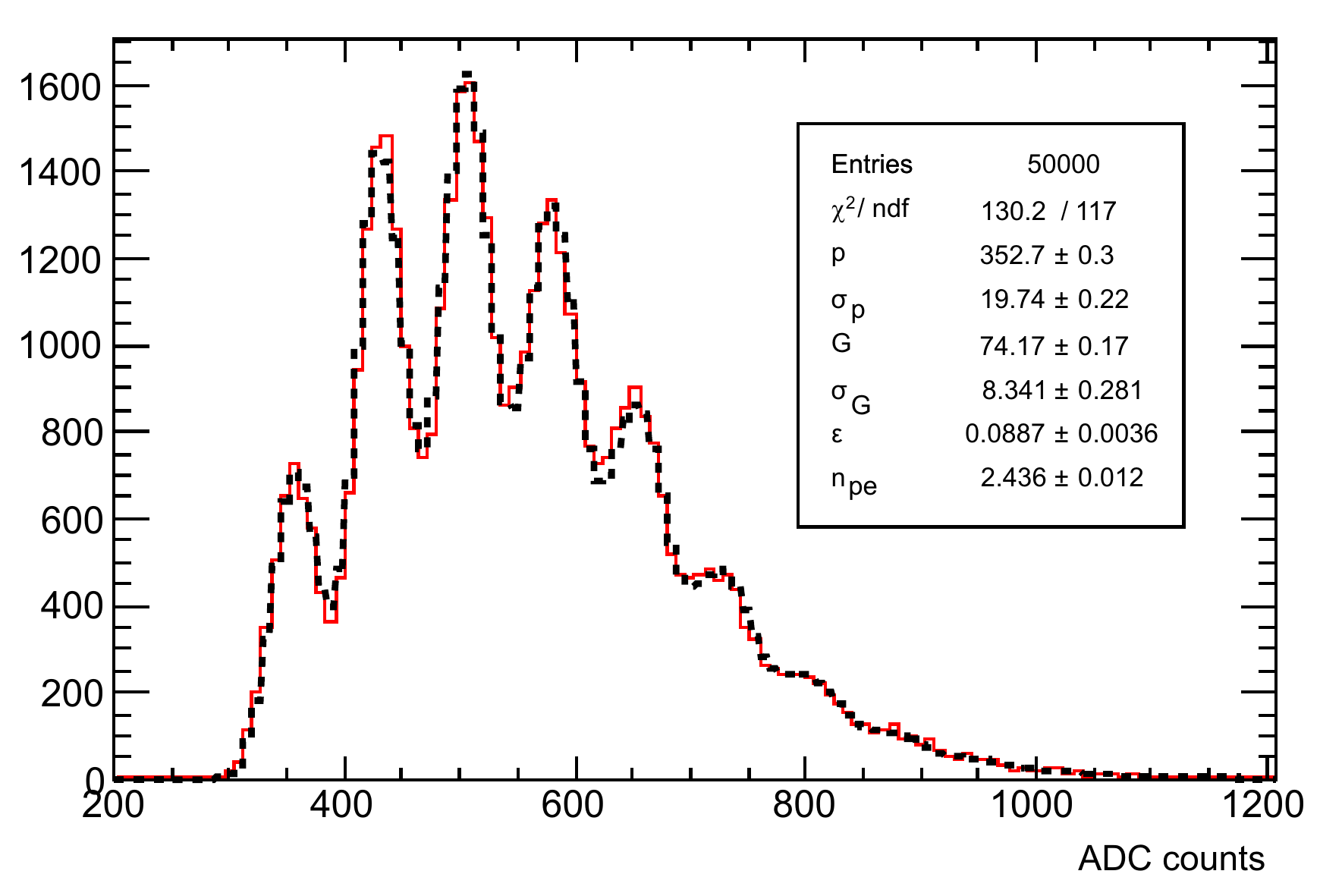}
\caption{An LED spectrum of a Hamamatsu MPPC 5883 (solid line) together with the fitted model
    curve (dashed line) used for the determination of the calibration
parameters (sec.~\ref{sec:led}).}
\label{fig:ledSpectrum}
\end{center}
\end{figure}
The measurements of the photon detection efficiency, homogeneity, and
crosstalk presented here rely on the analysis of LED spectra (fig.~\ref{fig:ledSpectrum})
which are fitted with the model of Balagura et al.~as described in~\cite{balagura}. In this
model, the five free parameters characterizing the SiPM are
the pedestal position $p$, the pedestal width $\sigma_p$,
the gain $G$, the signal width $\sigma_G$ describing the spread of signals
from different pixels, and the pixel crosstalk probability $\epsilon$
(sec.~\ref{sec:xtalk}).
In addition, the shape of the spectrum depends on the mean number of
photo-electrons $n_\mathrm{pe}$.
An example of the fit is given in figure~\ref{fig:ledSpectrum}.
The width of the $n$-th photo-electron peak is approximately given by
$\sigma_n=\sqrt{n\sigma_G^2+\sigma_p^2}$ so that only the first few
photo-electron peaks can usually be resolved. Here, the gain is simply
the spacing between two photo-electron peaks, measured in ADC counts.
Pixel crosstalk leads to a deviation from the underlying Poissonian
shape.

\subsection{Fill factor}
\label{sec:ff}
\begin{figure}
\begin{center}
\includegraphics[width=\columnwidth]{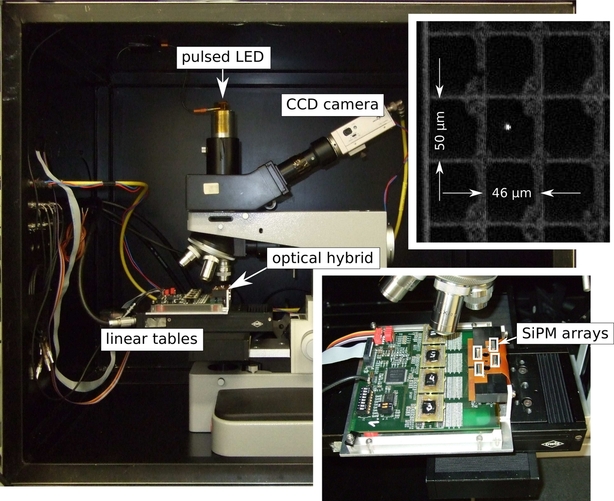}
\caption{Setup to determine the fill factor of SiPMs. {\it Top right:}
Close-up view of the LED focus achieved. The size of the focus should
be compared to the size of the SiPM pixels visible in the background. {\it Bottom right:} Close-up
view of an optical hybrid equipped with four SiPM arrays, together
with readout electronics, in the object plane of the microscope.}
\label{fig:fillFactorSetup}
\end{center}
\end{figure}

The ratio of sensitive to total area of an SiPM is called geometric efficiency or fill factor 
$\epsilon_\mathrm{FF}$. A setup to determine $\epsilon_\mathrm{FF}$ has been built in Aachen 
(fig.~\ref{fig:fillFactorSetup}).
It consists of a commercial microscope in which an LED is mounted 
behind a small pinhole. Using the light path of the microscope, the spot is focused down to less than 
$\unit[5]{\upmu{}m}$. The object plane can be moved two-dimensionally
with high precision linear tables using
computer-controlled stepper motors.

To determine $\epsilon_\mathrm{FF}$, an SiPM is mounted on the linear
tables.
The surface is scanned in both directions and the 
surface of the SiPM is illuminated by a series of $\unit[5]{ns}$
short LED flashes. We define that a point at which a pixel fires
at least \unit[95]{\%} of the time belongs to the sensitive area.
For the determination of the fill factor, a very fine grid of
$\unit[1]{\upmu{}m}$ spacing is used. By using a coarser scanning
grid, SiPM arrays can routinely and quickly be tested for dead
pixels before assembly of the optical hybrids.

\begin{figure}
\begin{center}
\includegraphics[width=\columnwidth]{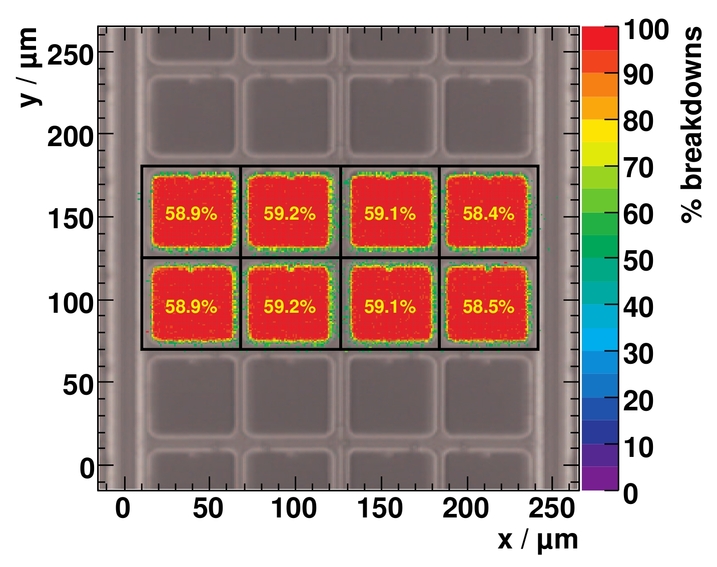}
\caption{Determination of the fill factor for eight pixels of an MPPC 5883,
using an extremely fine grid of sampling points with the microscope setup
shown in fig.~\ref{fig:fillFactorSetup}. A microscope picture of the SiPM surface is shown
in the background. The scale gives the percentage of breakdowns triggered at each scan
position.}
\label{fig:fillFactor}
\end{center}
\end{figure}

In fig.~\ref{fig:fillFactor}, the 
fill factor of eight pixels of an Hamamatsu MPPC 5883 array is
plotted as an example, together with a microscope picture of the device.
Averaging over the individual pixels yields a mean pixel fill factor
of $\epsilon^\mathrm{pix}_\mathrm{FF}=\unit[59.4]{\%}$. In the determination of
the overall photon detection efficiency however, the gaps between
neighboring SiPM channels have to be taken into account, and the fill
factor, averaged over the channel width of $\unit[250]{\upmu{}m}$,
then becomes $\epsilon_\mathrm{FF}=\unit[55.2]{\%}$.

\subsection{Photon Detection Efficiency}
\label{sec:pde}
The photon detection efficiency $\epsilon_\mathrm{PDE}$ is measured by
shooting LED flashes onto an SiPM and comparing the mean number of detected
photons to those of a calibrated photomultiplier tube with known quantum efficiency
$\epsilon_\mathrm{PMT}$. Considering the sensitive areas
$A_\mathrm{PMT}$ and $A_\mathrm{SiPM}$, as well as
the mean number of photo-electrons $\bar{n}_\mathrm{PMT}$ and $\bar{n}_\mathrm{SiPM}$ of the
PMT and SiPM, respectively, the photon detection efficiency can be determined from
\begin{equation}
\label{eq:pde}
\frac{\bar{n}_\mathrm{PMT}}{\epsilon_\mathrm{PMT} \cdot A_\mathrm{PMT}} = 
\frac{\bar{n}_\mathrm{SiPM}}{\epsilon_\mathrm{PDE} \cdot A_\mathrm{SiPM}}
\end{equation}

\begin{figure}
  \begin{center}
  \includegraphics[width=\columnwidth]{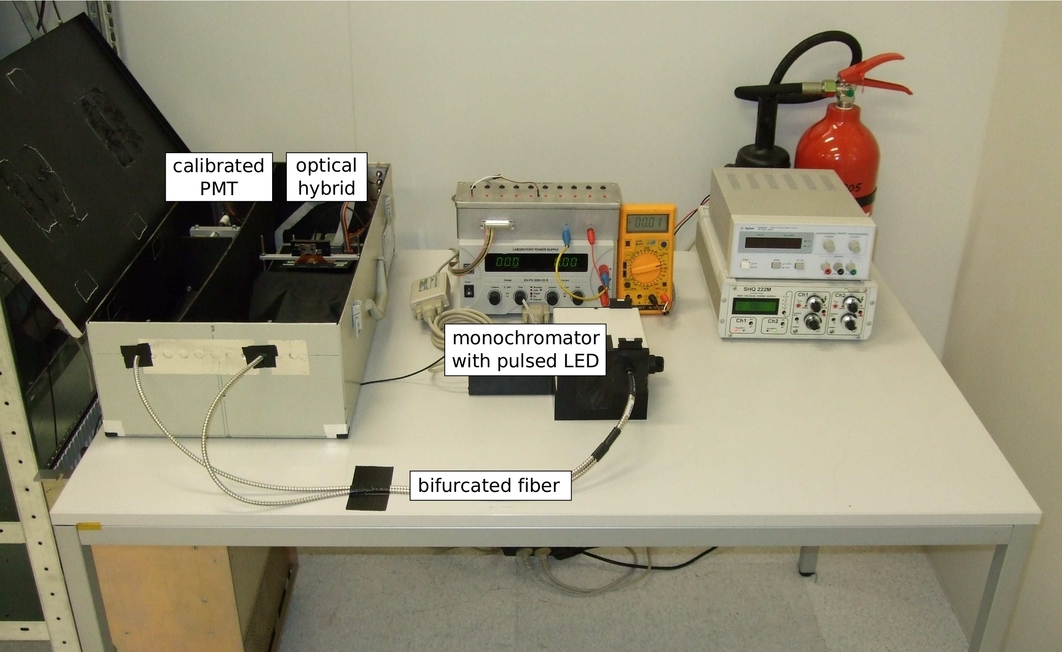}
  \caption{Setup to measure the photon detection efficiency of the SiPMs.}
  \label{fig:pdeSetup}
  \end{center}
\end{figure}

In the test setup (fig.~\ref{fig:pdeSetup}), the LED is flashed into a monochromator.
Its light is then split by a calibrated 50/50 bifurcated fiber and guided to two
optically separated sections of a light-tight box. The calibrated PMT and the optical
hybrid are placed at the same distance of $\unit[30]{cm}$. 
A diffusor is used to guarantee that the SiPM and the PMT are illuminated
uniformly. The response of the PMT is measured with a charge-to-digital
converter. The same model used to fit the SiPM spectra is employed to determine
the mean number of photo-electrons in the PMT by setting the pixel crosstalk
probability to zero.

$\epsilon_\mathrm{PDE}$ is measured as a function of the bias voltage $U$ applied
to the SiPM between \unit[70.0]{V} and \unit[71.75]{V} in \unit[0.25]{V} steps. The
power supply can be controlled from a lab PC, such that the whole characterization
is fully automatic. All measurements
have been performed at a temperature of $T=20^\circ\mathrm{C}$.

The photon detection efficiency depends on the overvoltage at which
the SiPM is operated, defined as $\Delta{}U=U-U_0$ where $U_0$ is the
breakdown voltage of the SiPM. $U_0$ is determined by measuring the
gain $G$ as a function of the bias voltage, at bias voltages $U>U_0$.
The resulting linear curve can then be extrapolated down to zero gain
and the corresponding voltage is $U_0$.

\begin{figure}
\begin{center}
\begin{tabular}{cc}
\includegraphics[width=0.48\columnwidth]{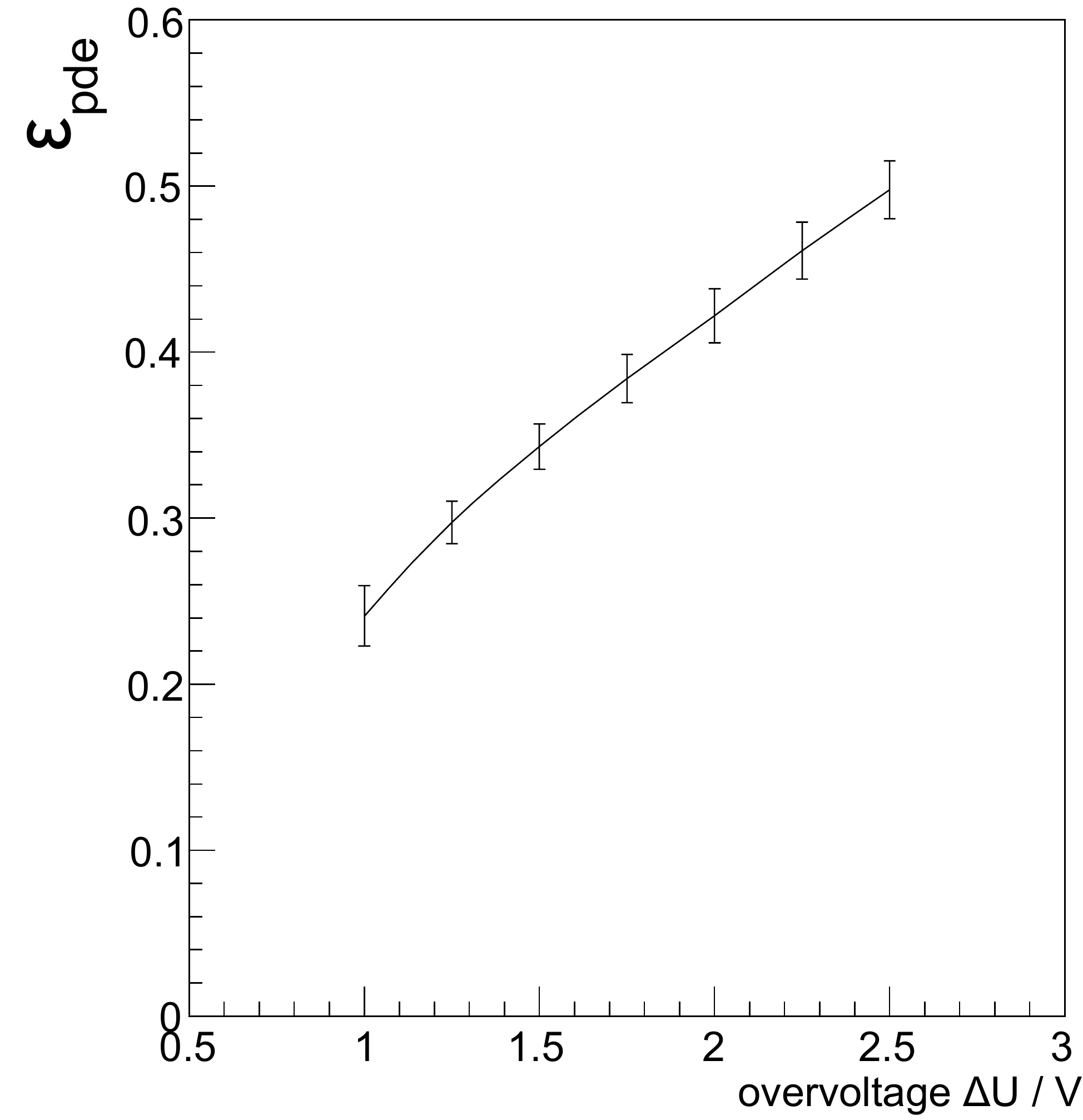}&
\includegraphics[width=0.41\columnwidth]{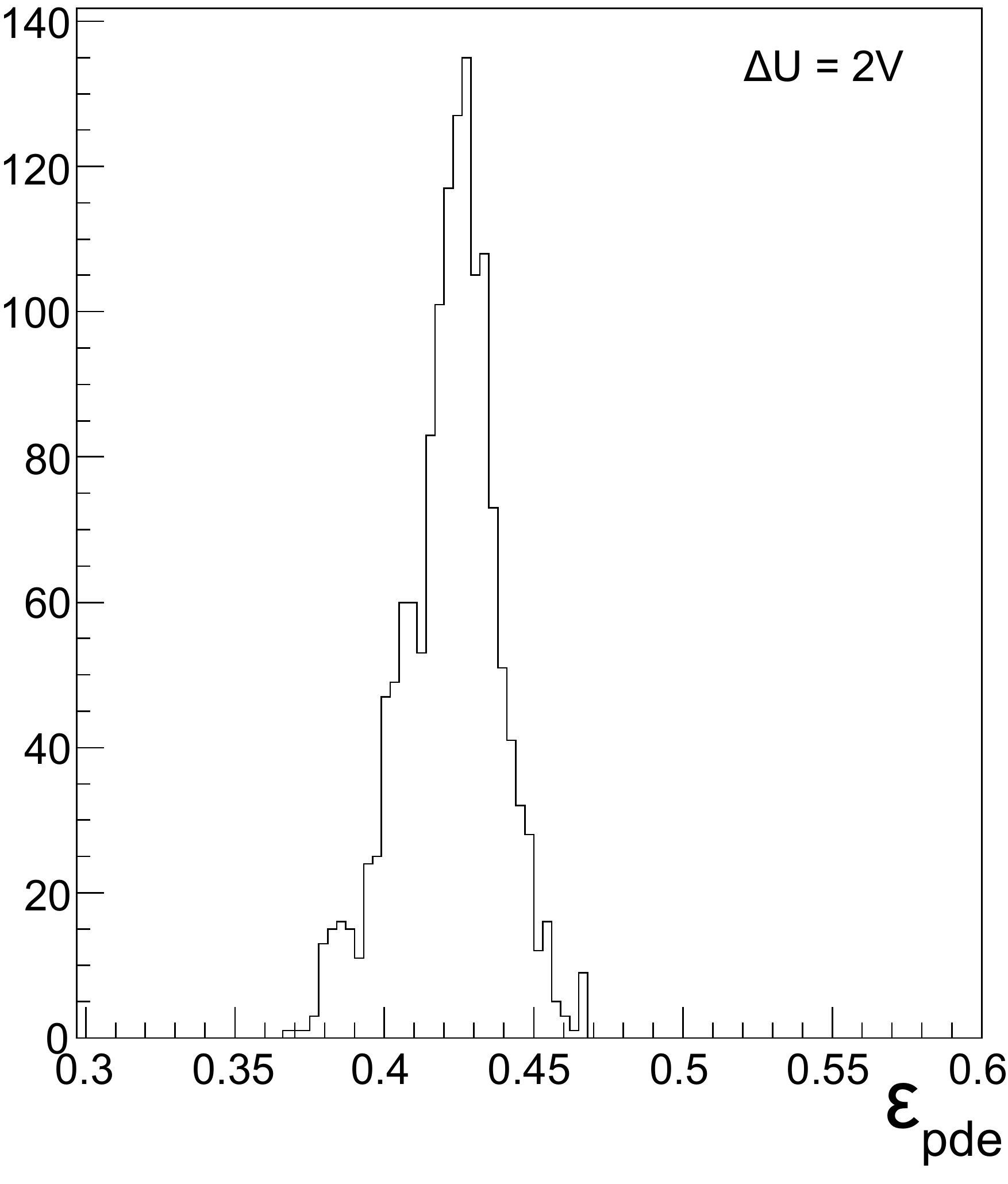}\\
\end{tabular}
\caption{Mean photon detection efficiency of 1441 channels of Hamamatsu MPPC 5883 devices as a
  function of the overvoltage at a wavelength of $\lambda=\unit[440]{nm}$. The line has been added
  to guide the eye (left).
  Photon detection efficiency distribution of 1441 channels on 46 SiPM arrays at a mean
  overvoltage of \unit[2.0]{V} (right).}
\label{fig:pdeVoltage}
\end{center}
\end{figure}

In fig.~\ref{fig:pdeVoltage} (left), the mean photon detection efficiency of 1441 channels
of Hamamatsu MPPC 5883 devices is plotted as a function of the mean overvoltage $\overline{\Delta{}U}=U-\bar{U}_0$,
at a fixed wavelength of
$\lambda=\unit[440]{nm}$. The error bars indicate the standard deviation of Gaussians fitted to the distribution of each point.
Here, $\bar{U}_0$ is the mean breakdown
voltage of the respective array.
As an example of the individual distributions, figure~\ref{fig:pdeVoltage} (right) contains
the values determined for $\epsilon_\mathrm{PDE}$ at an overvoltage of $\overline{\Delta{}U}=\unit[2.0]{V}$,
for 1441 channels on 46 arrays. 31 channels did not work because of broken electrical contacts. This problem
has been solved for the testbeam measurements by changing the connection technique from gluing to soldering.
The channel-to-channel variation is found to be at the level of a few percent, comparable to the variation in
breakdown voltage (sec.~\ref{sec:homo}).

$\epsilon_\mathrm{PDE}$ reaches a level of $50\,\%$ at
an overvoltage of $\unit[2.8]{V}$.
This result underlines that the silicon photomultiplier has reached a
level of maturity where its photon detection efficiency is mainly
determined by its geometrical properties, expressed in terms of the
fill factor, as comparison with eq.~(\ref{eq:epde}) shows that the
product $\epsilon_\mathrm{ABE}\cdot\epsilon_\mathrm{QE}$
must be close to unity. Nevertheless, care must be taken when
determining the optimal voltage at which to operate the SiPMs in the tracker. This
is because pixel and strip crosstalk increase along with $\Delta{}U$,
deteriorating the spatial information and thereby counteracting the
beneficial effect of the increased light yield.

\begin{figure}
\begin{center}
\includegraphics[width=\columnwidth]{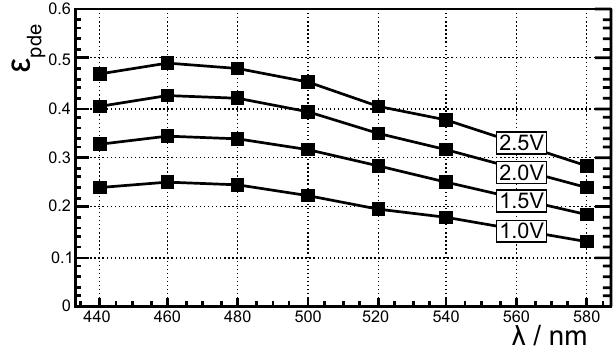}
\caption{Spectral response of a Hamamatsu MPPC 5883 for four
  overvoltages between $\unit[1]{V}$ and $\unit[2.5]{V}$. The photon detection
efficiency peaks at $\lambda=\unit[460]{nm}$. Lines have been added to guide the eye.}
\label{fig:spectralResponse}
\end{center}
\end{figure}

In fig.~\ref{fig:spectralResponse}, the spectral response of the
devices can be seen for four overvoltages between $\unit[1]{V}$ and $\unit[2.8]{V}$. The
peak sensitivity is reached at a wavelength of \unit[460]{nm} matching the peak emission wavelength of
Kuraray SCSF-81MJ scintillating fibers.

\subsection{Homogeneity}
\label{sec:homo}
\begin{figure}
\begin{center}
\begin{tabular}{cc}
\includegraphics[width=0.45\columnwidth]{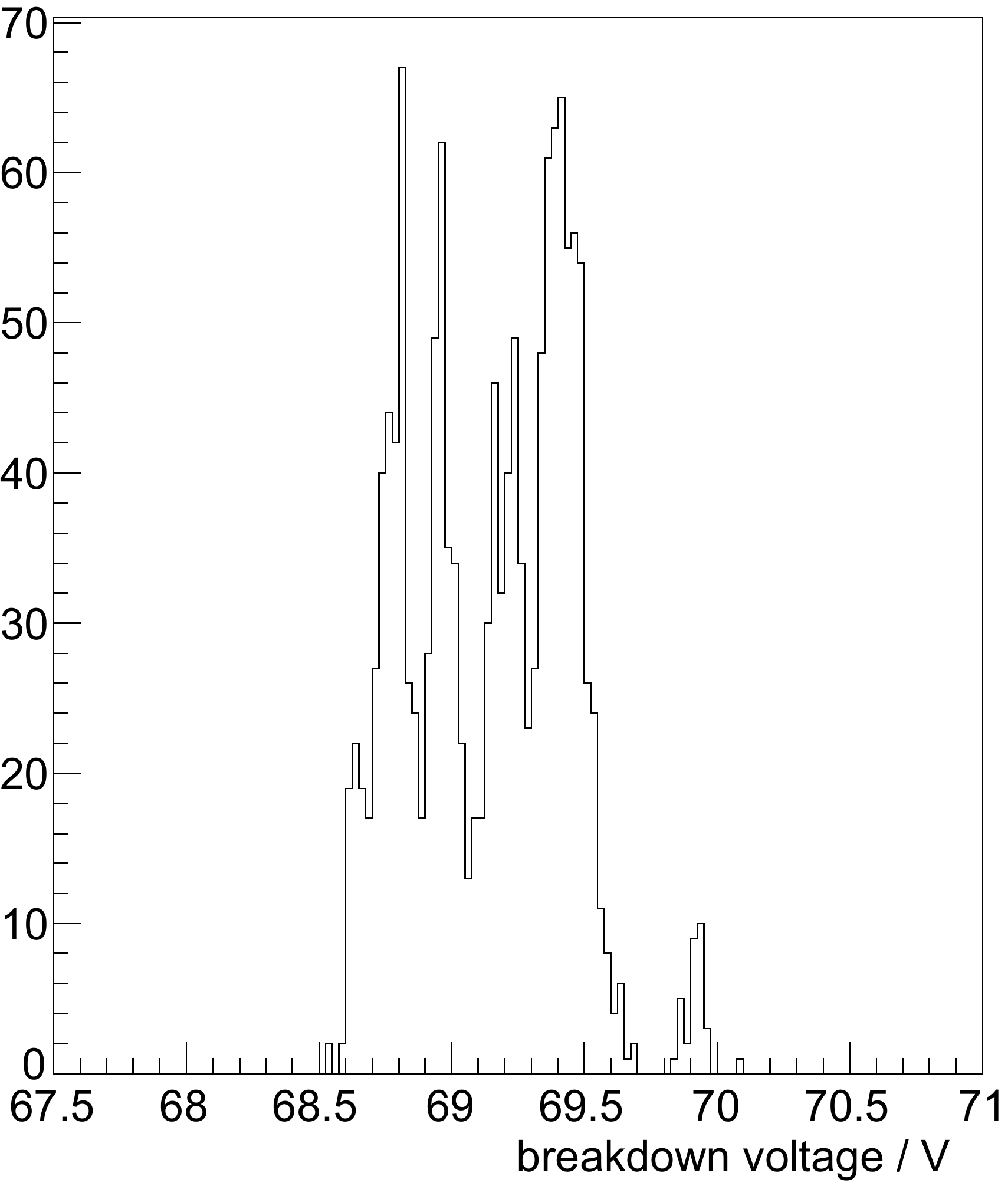}&
\includegraphics[width=0.45\columnwidth]{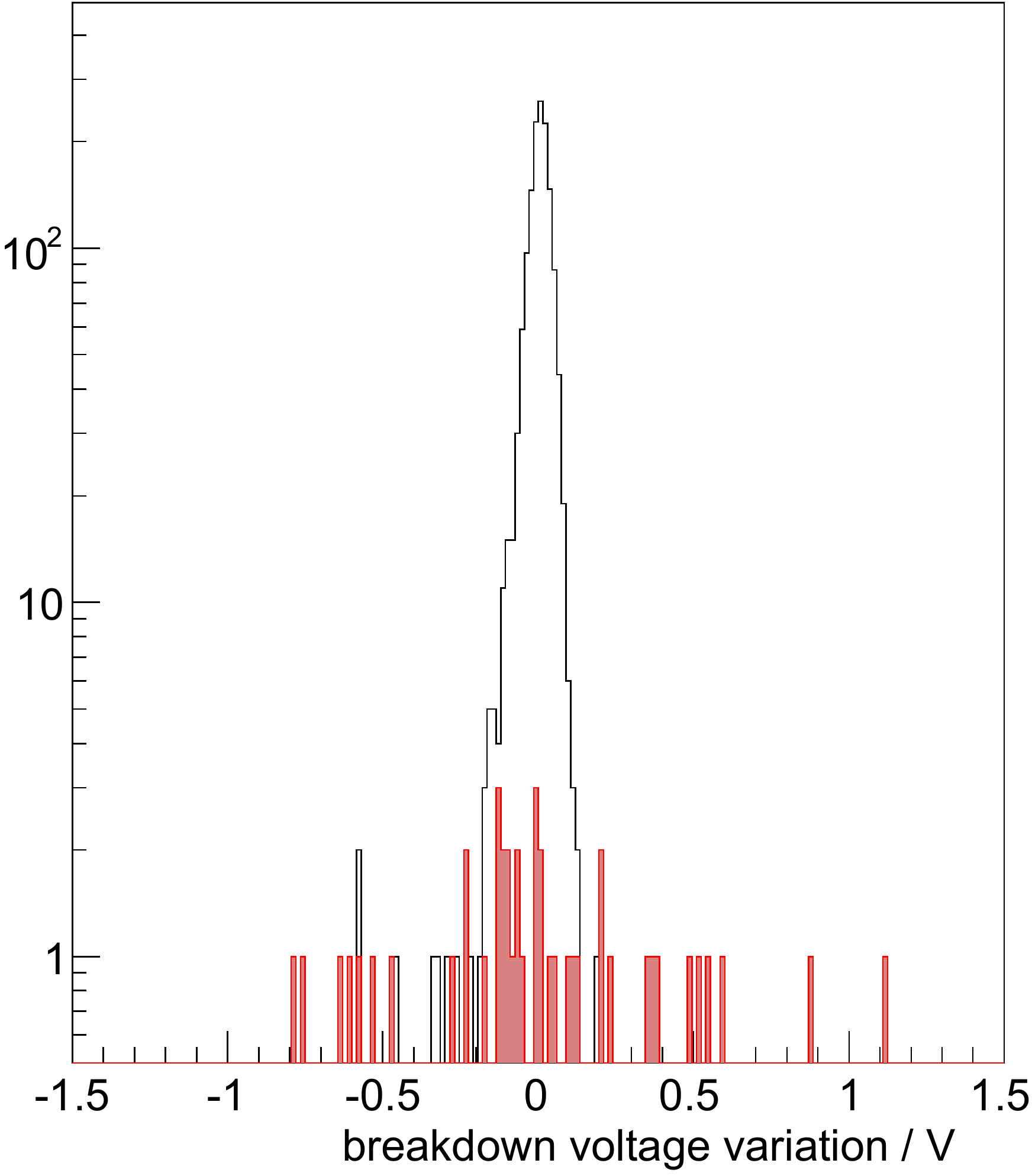}\\
\end{tabular}
\caption{Distribution of the breakdown voltage $U_0$ of 1441 channels of Hamamatsu MPPC 5883
  devices (left). After subtraction of the mean breakdown voltage of the
  corresponding array, the small channel-to-channel variation remains
  (right, unshaded histogram).
  The distribution obtained for the first channel in each array (shaded
  histogram) shows a larger variation.}
\label{fig:bdv}
\end{center}
\end{figure}

To study the homogeneity of the SiPMs, the breakdown voltages of
1441 channels on all 46 Hamamatsu MPPC 5883 arrays available in Aachen
have been determined (fig.~\ref{fig:bdv} (left)).
While the breakdown voltage varies by around $\unit[2]{V}$, we find an
RMS variation of the breakdown voltage from channel to channel of only $\unit[0.037]{V}$.
This is illustrated in fig.~\ref{fig:bdv} (right) showing the distribution of
$U_0-\bar{U}_0$. The breakdown voltages of the first channels in the
arrays scatter by a much larger amount. The cause for this effect is
currently under investigation.

\begin{figure}
\begin{center}
\includegraphics[width=\columnwidth]{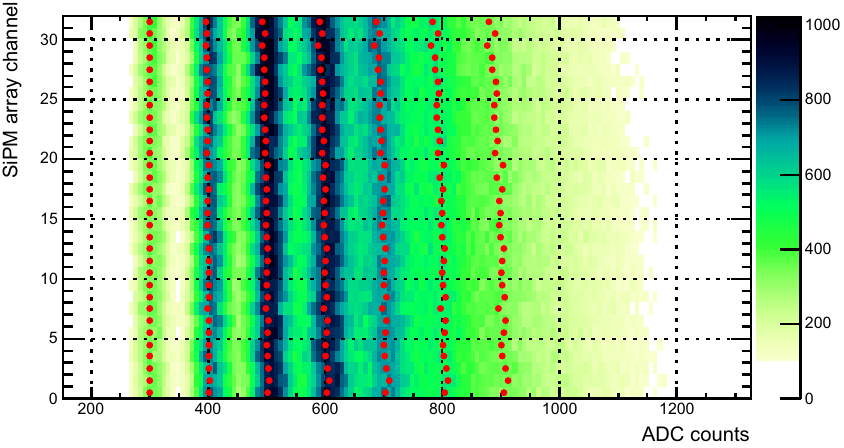}
\caption{To study the homogeneity of the SiPM response, LED spectra of the Hamamatsu MPPC 5883 devices
are measured. Each bin along the $y$-axis corresponds to one SiPM in the array. The pedestal
values have been equalized for better comparison of the gain values.
The dots mark the positions of the first few photo-electron peaks.}
\label{fig:ledSpectra}
\end{center}
\end{figure}

The homogeneity of the gain is demonstrated in figure~\ref{fig:ledSpectra}
showing LED spectra taken with a typical Hamamatsu MPPC 5883. The response
of each of the 32 channels, operated at the same bias voltage, is shown
after shifting the pedestal to the same value of 300 ADC counts. The
individual photo peaks are visible, illustrating the auto-calibration capability
of the SiPM. The gain $G$ is related to the distance between two peaks and is
clearly seen to be homogeneous over the entire array.

To summarize our findings, the key properties gain and photon detection efficiency were found
to be nearly identical at a fixed temperature for all SiPM channels of
all examined arrays as long as they are
operated at the same overvoltage. For a reliable operation of the tracker, $\Delta{}U$
therefore needs to be kept constant at an optimal value, and it has to be adjusted separately for each
array, but not for each channel, which greatly simplifies the operation.

\subsection{Crosstalk and noise}
\label{sec:xtalk}
\begin{figure}
\begin{center}
\includegraphics[width=\columnwidth]{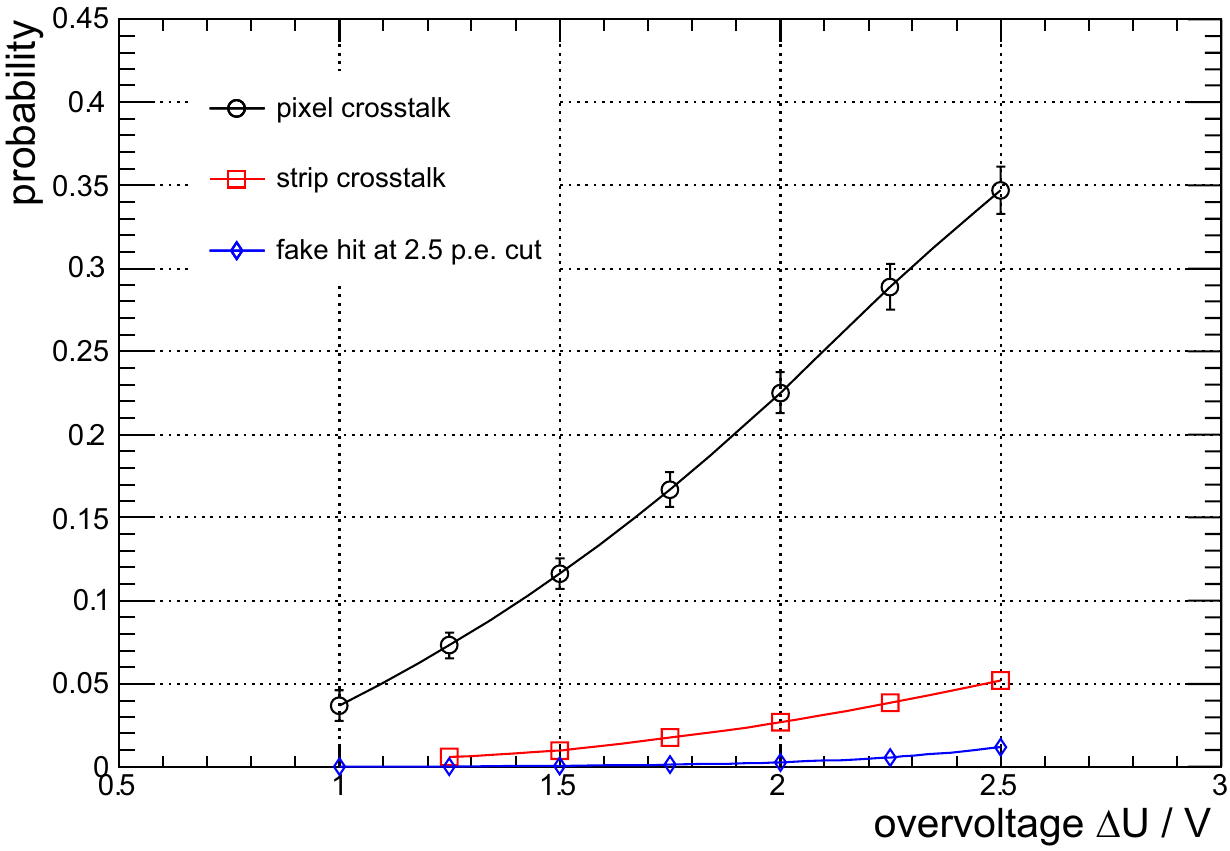}
\caption{Pixel crosstalk, strip crosstalk and fake hit probability of the SiPM arrays
as a function of the overvoltage. Lines have been added to guide the eye.}
\label{fig:allNoise}
\end{center}
\end{figure}

When an avalanche develops in a given pixel, photons are produced that
can subsequently trigger a
neighboring pixel. This effect is known as pixel crosstalk~\cite{ref:xtalk} and the
probability for a pixel to trigger a neighboring pixel is called the
crosstalk probability $\epsilon$.
It can be measured from the same LED spectra used for the
determination of the photon detection efficiency, as described in the beginning of
this section.
Figure~\ref{fig:allNoise} shows the pixel crosstalk
probability as a function of the overvoltage. Each point
shows the mean value of the examined sample and the error bars give
the root mean square variation across the sample.

Worse than pixel crosstalk, photons from the avalanche process can reach neighboring array strips and
cause them to fire. This effect is called strip crosstalk. The strip crosstalk probability $\epsilon_s$
is defined as the probability for any of the neighboring channels to be triggered by an SiPM breakdown. It can be
determined from noise
measurements by looking at coincidences of fired channels and correcting for
random coincidences. $\epsilon_s$ is found to be lower than $\epsilon$ by
roughly an order of magnitude (fig.~\ref{fig:allNoise}).
The crosstalk probabilities measured constitute a major problem in the
operation of the SiPM arrays, and these results provide key input to Monte Carlo studies used to
determine the ideal operating point in terms of the overvoltage.

When operated above $U_0$, SiPM pixels fire randomly
due to thermal generation of charge carriers, thus creating a fake hit
in the tracker module. Dark spectra have been
taken along with the LED spectra. The gain can be determined from the LED spectra and thus
a 2.5 photo-electron threshold can be applied to the noise spectra from which the
corresponding fake hit probability per array strip can be determined
(fig.~\ref{fig:allNoise}).
At $T=20^\circ\mathrm{C}$, a probability of
\unit[0.3]{\%} is reached at \unit[2.0]{V} overvoltage.
A scintillating fiber tracker for use in particle or astroparticle physics will
typically consist of $\mathcal{O}(4-12)$ layers of fiber
modules. Track finding algorithms to identify the hits belonging to a
particle trajectory in the presence of fake hits have been
implemented~\cite{ref:henning_diss} that can easily cope with this kind of
noise level.

\section{Testbeam measurements 2009}
\label{sec:testbeam}
\subsection{Setup description}
\label{sec:testbeamsetup}
\begin{figure}[t]
\begin{center}
\begin{tabular}{cc}
\includegraphics[height=0.45\columnwidth,angle=0]{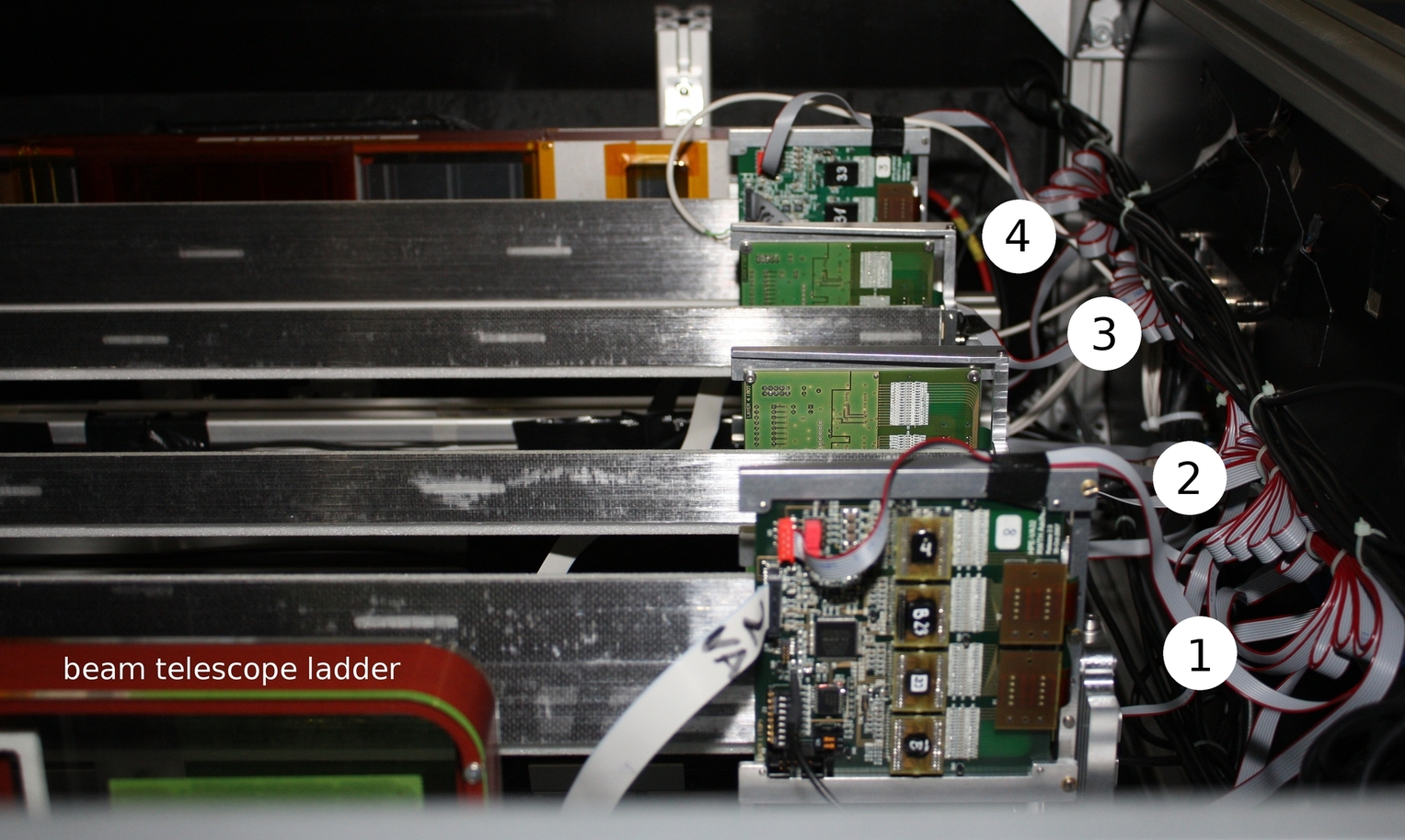}&
\includegraphics[height=0.45\columnwidth,angle=0]{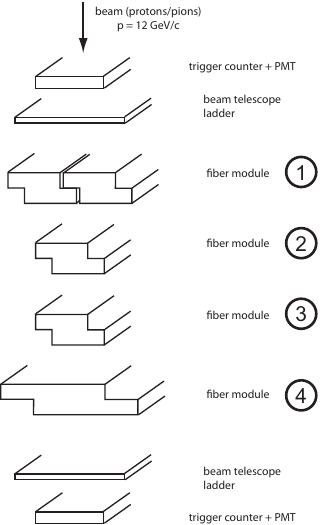}\\
\end{tabular}
\end{center}
\caption{A perspective photograph (left) and schematic drawing (right)
of the 2009~testbeam setup. The five modules studied in the testbeam
are visible: two narrow modules (128~fibers wide, (2) and (3), fig.~\ref{fig:tb09_module}), one
wide module (256~fibers wide, (4), and two narrow modules mounted
next to each other (1), with the aim of studying the small gap between
the modules. The mounting frames of the silicon beam telescope ladders
are visible in front of and behind the four layers of fiber modules,
respectively. The overall dimensions of the testbeam box were
$120\times120\,\mathrm{cm}^2$.}
\label{fig:tb09_photo}
\end{figure}
\begin{figure}[htb]
\begin{center}
\includegraphics[width=\columnwidth,angle=0]{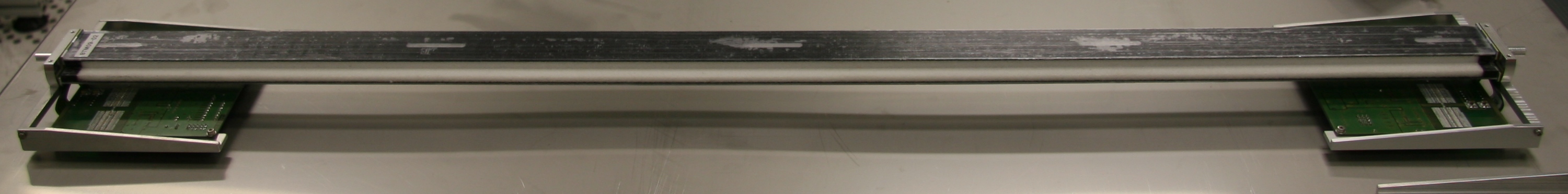}
\end{center}
\caption{Photograph of a completed narrow module used in the
testbeam. Two boards carrying front-end electronics are visible at the sides.}
\label{fig:tb09_module}
\end{figure}
\begin{figure}[htb]
\begin{center}
\includegraphics[width=0.8\columnwidth,angle=0]{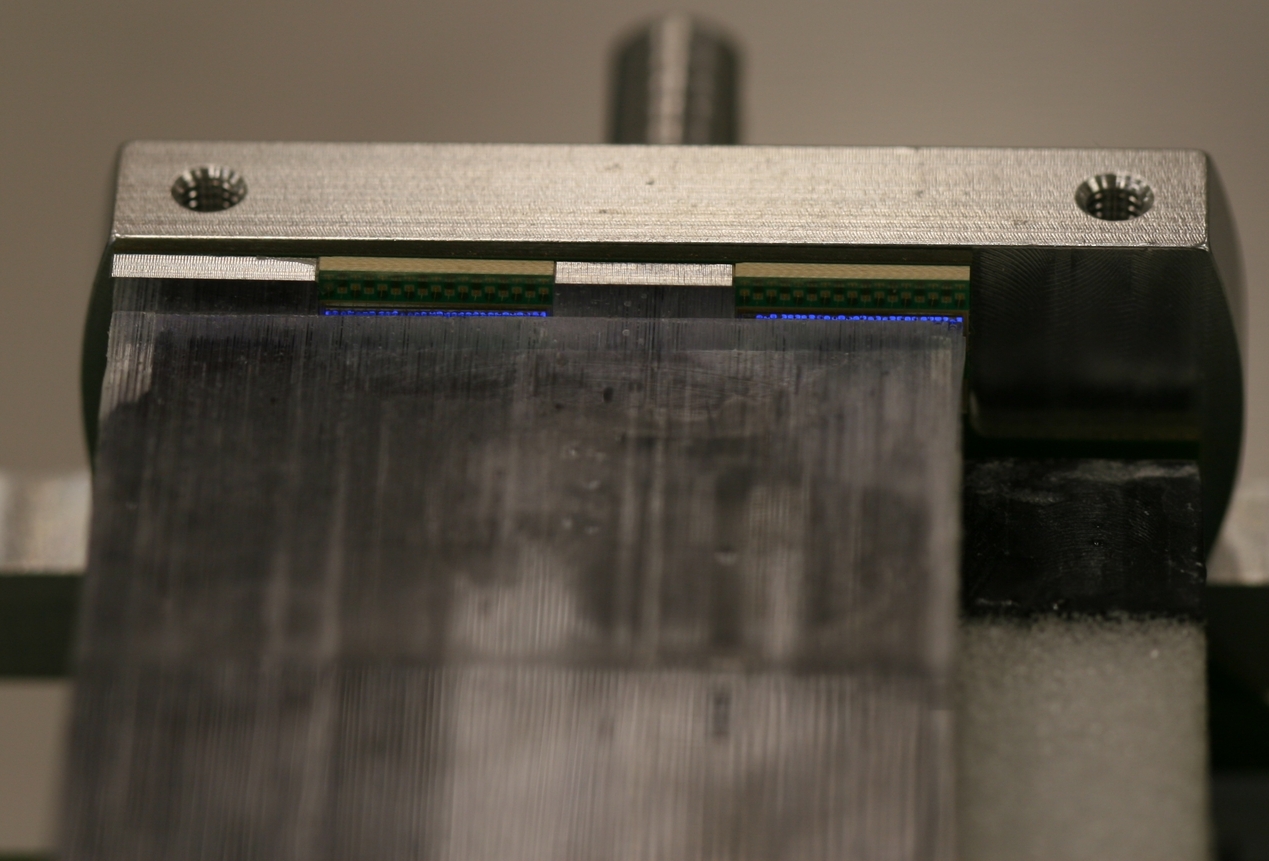}
\end{center}
\caption{Detailed view of the mounting of a hybrid carrying a silicon photomultiplier
array between the end of a fiber ribbon and an aluminum carrier
piece. Two mirrors are visible next to two SiPM arrays.}
\label{fig:tb09_moduleEnd}
\end{figure}
Five $\unit[860]{mm}$ long prototype modules for the fiber tracker have been
subjected to a testbeam at the CERN T9 beam-line in
fall~2009. The beam consisted mainly of pions and protons and had a
defined momentum of $p=\unit[12]{GeV}/c$. The main goals were the determination of the spatial
resolution and of the light yield of the modules.

The testbeam setup (figs.~\ref{fig:tb09_photo},
\ref{fig:tb09_module}, and~\ref{fig:tb09_moduleEnd})
consisted of two trigger scintillator counters of dimension $20\times10\times1\,\mathrm{cm}^3$, two
double-sided silicon-strip detectors built for the AMS-02
tracker~\cite{alcaraz08}
and the five fiber module prototypes that were arranged in four layers
(cf.~fig.~\ref{fig:tb09_photo}).
To enable comparison studies, four of the five fiber modules were made
of Kuraray SCSF-78MJ fibers while the remaining one was made of
SCSF-81MJ fibers. The SCSF-78MJ fibers became available just in time
for the testbeam. In addition, for three optical hybrids (12~SiPM
arrays), optical grease of type NyoGel OCK-451 or OC-459~\cite{ref:grease}
was applied to the fiber-SiPM
interface. As the beam spot size (rms width) was only about
$\unit[8]{mm}$ in each direction, not all SiPM arrays were illuminated
and could be included in the study.

The silicon-strip
detectors were used as a beam telescope, allowing a reference
measurement of the trajectory of a beam particle which is vital for
the determination of the spatial resolution.
Under ideal conditions, the beam telescope ladders themselves each provide a spatial resolution of approximately
$\unit[10]{\upmu{}m}$ and $\unit[30]{\upmu{}m}$ in the
coordinates perpendicular and parallel to the fiber axis,
respectively~\cite{bt_res}. The corresponding readout pitch is
$\unit[110]{\upmu{}m}$ and $\unit[208]{\upmu{}m}$ for the strips
parallel and perpendicular to the fibers, respectively.
The trigger scintillators were connected to
ordinary PMTs.
For the readout of the SiPM arrays, two different preliminary
electrical hybrid boards were available, one based on
the IDEAS VA\_32/75 readout chip~\cite{ideas} and the other based on
the SPIROC~\cite{ref:spiroc} readout chip, specifically designed for
the readout of SiPMs, with 32 channels each.
The temperature at which the SiPMs were operated varied between $22^\circ\mathrm{C}$ and
$25^\circ\mathrm{C}$ depending on location in the setup and the
day-night cycle. However, thanks to the excellent properties of the
used SiPMs, noise did not pose a problem even at these high
temperatures.

The trigger scheme was based on NIM electronics and
had to be optimized for speed as the hold
signal for the VA\_32/75 and SPIROC chips reading out the SiPMs had to arrive
around $\unit[85]{ns}$ after the particle crossing as the SiPM
signal pulse reaches its maximum at that time. This was achieved by
using only a single coincidence of the two PMTs reading out the
trigger scintillators. A post-event dead-time of $\unit[150]{\upmu{}s}$ was started
along with the hold signal, and a pre-event dead-time of
$\unit[5]{\upmu{}s}$ was started by a signal from any of the trigger
PMTs. Upon arrival of a trigger signal, the raw data (ADC counts) of
all readout channels present in the setup were stored on disk for
offline analysis.

The coordinate system was chosen such that the beam traversed the
setup in the $z$-direction, the fibers were parallel to the
$y$-direction and therefore the fiber modules measured the
$x$-direction of a particle track. The double-sided silicon beam
telescope modules measured both in the $x$- and $y$-directions.

\subsection{Analysis procedure}
\label{sec:analysis}
The analysis procedure begins with the calibration of the silicon beam
telescope ladders and the SiPM arrays in front of the fiber
modules. For this purpose, dedicated calibration runs
were taken between spills, before each data run. In a first step, dark events were
taken. For the ladders, the pedestal and noise for each strip are
determined as the mean and root mean square of the distribution of raw
ADC counts. For the SiPM arrays, the pedestals for all channels are
determined accordingly. In the second step, LEDs located next to each
fiber module were pulsed with a pulse length of $\unit[6]{ns}$
and the SiPM parameters, especially gain and
crosstalk, are determined from fits to the spectra, as described in
sec.~\ref{sec:led}. Here, the uniformity of the SiPM response across
an array, as demonstrated above, is exploited to speed up the
calculations by fitting only the averaged spectrum.

Using the calibration parameters so obtained, clusters can be identified
-- neighboring channels with amplitudes significantly above the
background caused by noise. For the beam telescope ladders, a seed cluster
is formed when a strip with a significance exceeding $5\sigma$ is
found. Neighboring strips are added to the cluster as long as their
significance is at least $2\sigma$. If no neighboring cluster
fulfilling this condition is found, the one with the highest
significance is added nevertheless, in order to make use of the
charge-sharing between adjacent strips which improves the spatial
resolution~\cite{ref:azzarello}.\\
For the SiPM arrays, the raw SiPM data has to be converted to the number of fired pixels.
With the pedestal position $p_i$ for a given SiPM channel $i$ and the
gain $g$, as determined during the calibration and constant across an
array, the number of pixels
$s_i$ for a given channel is then simply calculated from the raw ADC
amplitude $a_i$ according to
\begin{equation}
\label{eq:sipmcalibration}
s_i=\frac{a_i-p_i}{g}
\end{equation}
Then, fiber clusters are
identified, as the photon output of the scintillating fibers for a single minimum
ionizing particle is spread over several neighboring SiPM array channels. Starting from a channel with at least
$2.5$~photo-electrons, neighboring channels are added to the cluster
as long as
their amplitude exceeds $0.5$~photo-electrons. The position of the
fiber cluster is then calculated as
\begin{equation}
\label{eq:clusterwmean09}
x_\mathrm{cl}=\frac{\sum\limits_{i}s_ix_i}{\sum\limits_{i}s_i}
\end{equation}
where the sum is taken only over the channel with the highest
amplitude and its immediate neighbors at most. We found that this
prescription slightly improves the spatial resolution.

Each cluster constitutes a potential intersection point of a particle
track. In the next step of the analysis chain, we therefore identify straight
line tracks, using an algorithm that is based loosely on the Hough
transform as implemented in~\cite{ref:hough}. It exploits the fact that each pair of
clusters belonging to the same track will yield similar values for the
slope and intercept of the line traversing the two clusters. To
determine the track parameters, we perform two simple straight line
fits in the $xz$- and $yz$-planes, respectively.

Using the tracks and clusters of the complete event sample, we then
performed a detector alignment using the Millepede
code~\cite{ref:millepede}. We allowed for deviations of the positions
of all the SiPM arrays from the nominal ones in the $x$-direction. The
Millepede algorithm minimizes the
global $\chi^2$ between tracks and clusters and yields unbiased
alignment parameters which we then used for a second iteration of the
analysis. We note that possible tilt angles of the fiber modules with
respect to each other and the beam telescope modules can be neglected
because of the small beam spot size and a mounting precision of better
than $\unit[2]{mrad}$. In addition, due to the very small angular
spread of the beam, a possible deviation from the nominal position along
$z$ would simply be translated into an apparent shift along $x$.

To identify clean single-track events, a number of quality cuts are
imposed. Fake tracks, identified by an anomalously high $\chi^2$ are
eliminated. Only clean beam telescope tracks, defined as having
exactly one cluster on each side of each ladder, are allowed. In
addition, tracks containing beam telescope clusters with too many
($\geq6$) strips or unusually low or high amplitude, indicative of a
dead or noisy strip, are removed. To further suppress spurious tracks,
the good collimation of the beam was exploited by doing appropriate
cuts in the $(m_x,x_0)$- and $(m_y,y_0)$-planes, where $m$ denotes the
reconstructed slope of a particle track and $x_0$ denotes its
offset. In total, we used roughly $60\,000$ clean single-track events
for this analysis.

As an important cross-check, we reconstructed the position
corresponding to each SiPM readout channel by using the beam telescope
information alone and histogramming the interpolated track position if
a given SiPM channel showed an amplitude exceeding
3.5~photo-electrons. The average distance between the mean values of
these histograms is then the reconstructed readout pitch. It was found
to agree with the nominal value of $\unit[250]{\upmu{}m}$ to within
$\unit[2]{\upmu{}m}$.

\subsection{Results}
\label{sec:testbeam_results}
The key figures of merit of a complete tracker module are the spatial
resolution and the tracking efficiency that it offers. 
These in turn depend crucially on the
overall light yield, expressed as the number of primary
photo-electrons (number of incident photons that triggered a Geiger
discharge in one of the pixels)
detected by the SiPMs, achieved for a minimum ionizing particle
crossing the module perpendicularly. 

\begin{figure}[htb]
\begin{center}
\begin{tabular}{cc}
\includegraphics[width=0.45\columnwidth,angle=0]{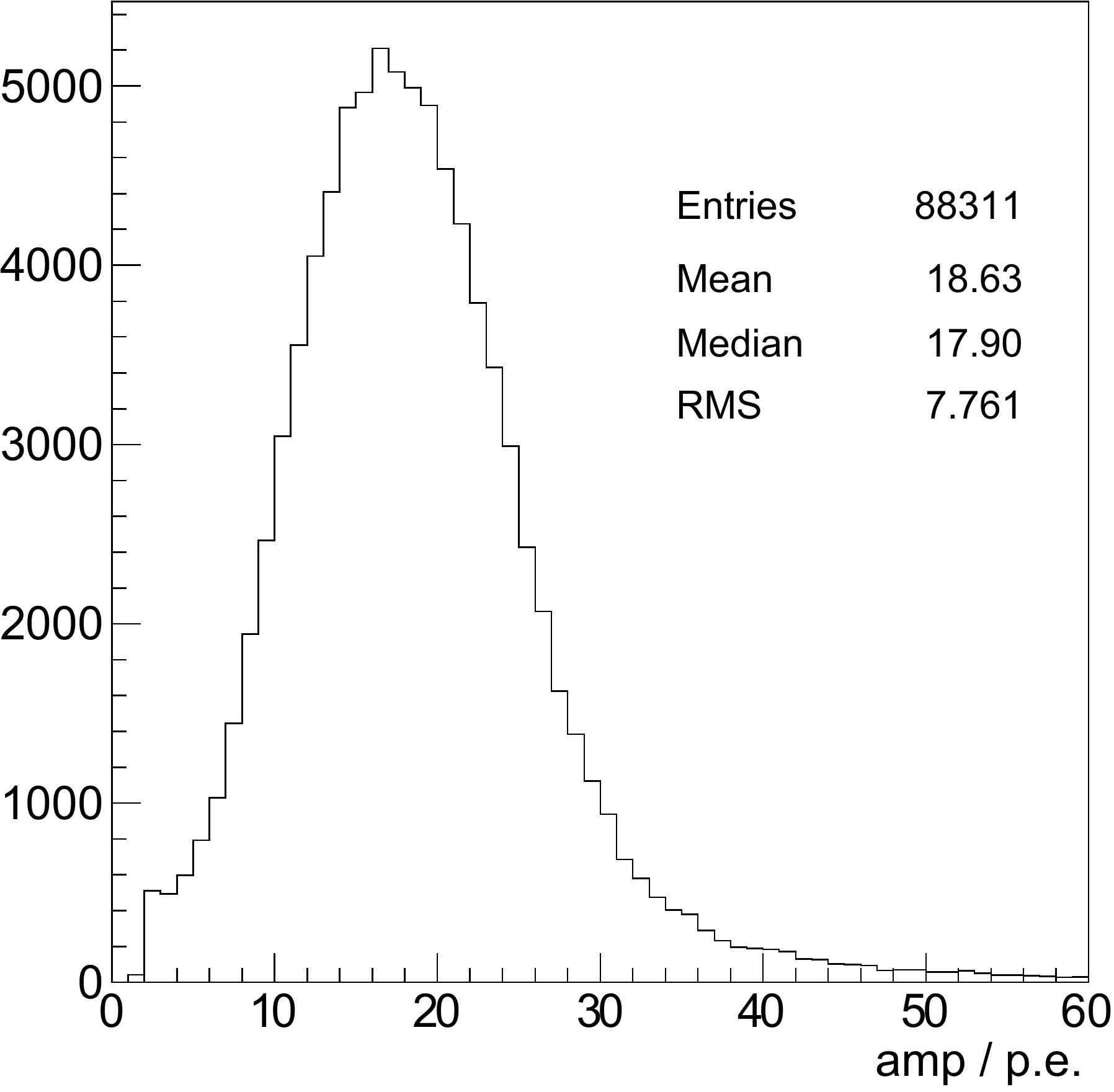}&
\includegraphics[width=0.42\columnwidth,angle=0]{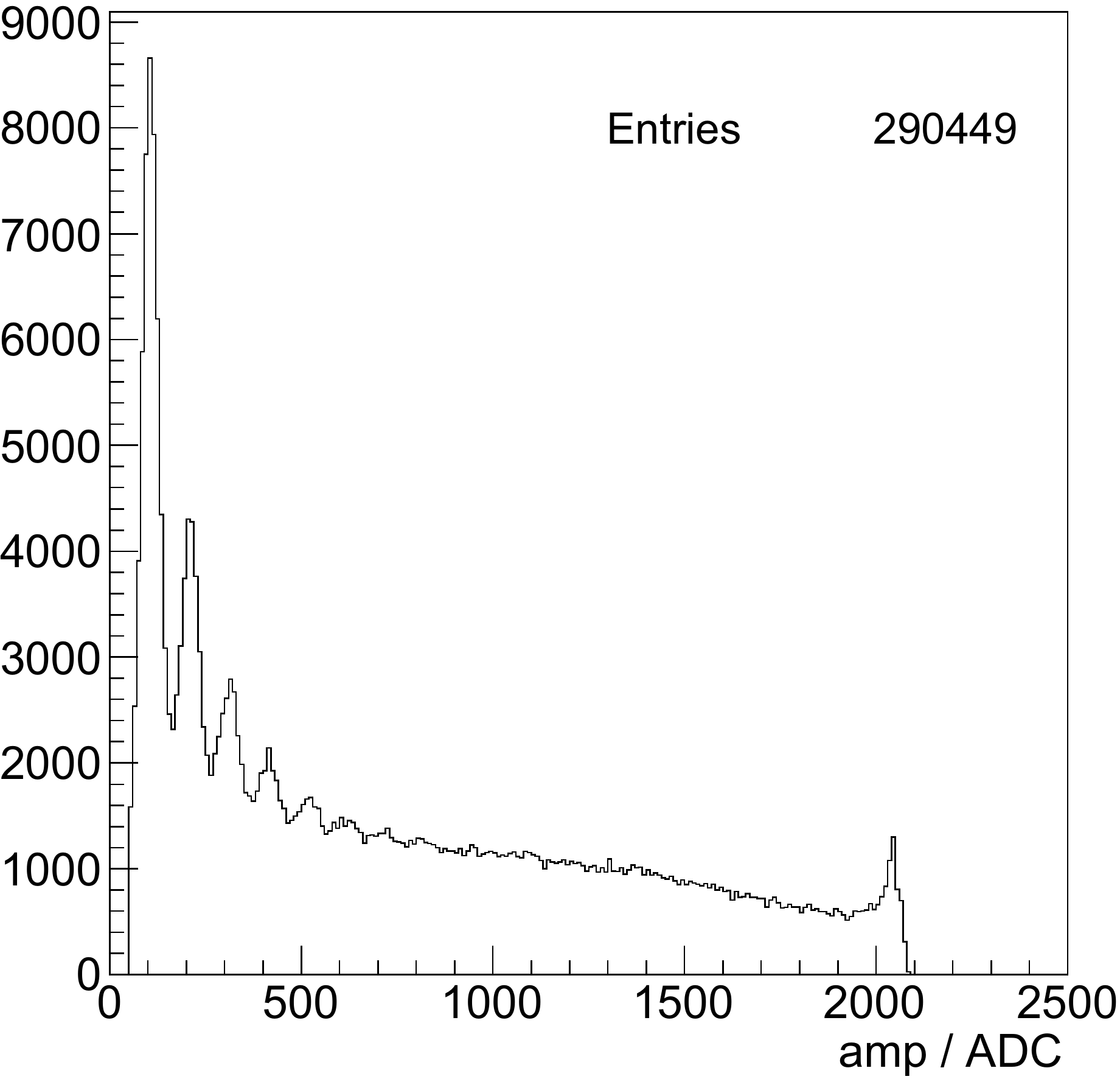}\\
\end{tabular}
\end{center}
\caption{Distribution of cluster amplitudes, after correction
for crosstalk and saturation, for a typical SiPM array
(left). Distribution of individual strip amplitudes in each cluster,
for the same array (right).}
\label{fig:tb09_amps}
\end{figure}

With the identification of fiber clusters as outlined in
the previous section, the raw cluster amplitude -- expressed as the
total number of fired pixels -- is easily
calculated. To obtain the corresponding number of photo-electrons, this
value has to be corrected for saturation effects and for crosstalk.
Saturation effects occur due to the limited number of SiPM
pixels so that two photons will occasionally hit the same pixel. The
situation is analogous to the combinatorial problem of distributing
$N$ balls into $M$ urns, and the necessary correction is found to
be~\cite{ref:urnmodel}
\begin{equation}
\label{eq:pixelcorr}
N_\mathrm{p.e.}=\frac{\log\left(1-\frac{N_H}{N_\mathrm{pix}}\right)}{\log\left(1-\frac{1}{N_\mathrm{pix}}\right)}\,\cdot\,(1-\epsilon)
\end{equation}
where $N_\mathrm{pix}=80$ is the number of pixels per SiPM, $N_H$ is the
number of fired pixels and $N_\mathrm{p.e.}$ is the calculated number
of photo-electrons. The factor $(1-\epsilon)$ in eq.~(\ref{eq:pixelcorr})
accounts for crosstalk~\cite{balagura}. The
combined probability for crosstalk was determined
from the LED spectra taken during the testbeam and was typically of
the order $\epsilon\sim0.25$.
The correction is applied on a per-channel
basis. The distribution of the
cluster amplitudes corrected for crosstalk and saturation effects is
shown in figure~\ref{fig:tb09_amps} for a typical SiPM array.
While the exact shape of the distribution is governed by many factors,
namely the fluctuations in primary energy deposition, fluctuations in
the path-length of the primary particle in the fiber ribbon, Poisson
fluctuations in the number of detected photons, and the cluster
definition used, its median is by definition taken to be the light yield
of the fiber-SiPM compound.
Figure~\ref{fig:tb09_amps} also contains the raw amplitude
distribution (after pedestal subtraction) of all the strips in
clusters on the same hybrid. Here, the individual photo-electron peaks
are well separated, demonstrating the auto-calibration capabilities of
the SiPM arrays. The small saturation bump towards high ADC counts is
due to the limited dynamic range of the VA readout chips used.

\begin{figure}[htb]
\begin{center}
\begin{tabular}{cc}
\includegraphics[width=0.45\columnwidth,angle=0]{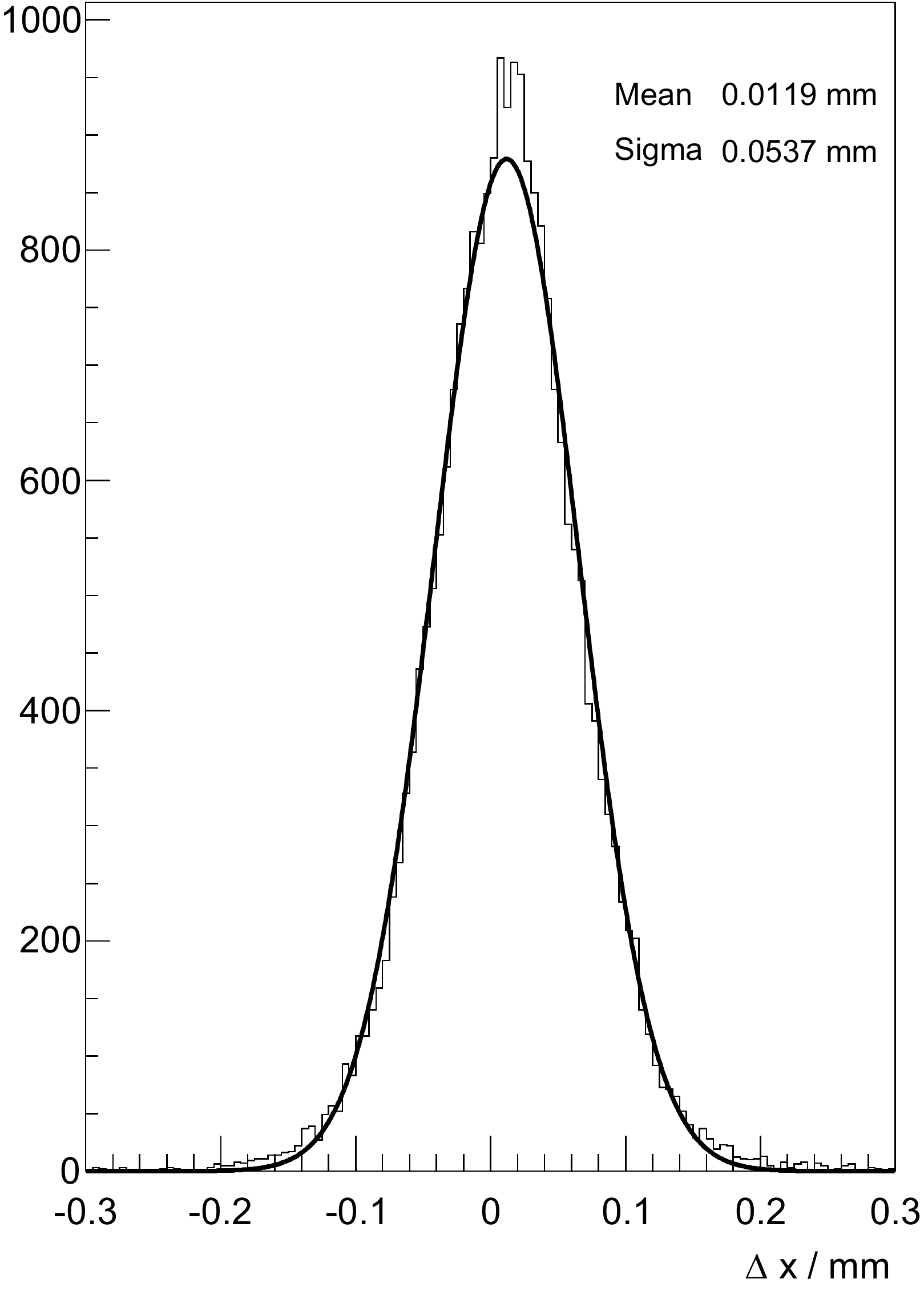}&
\includegraphics[width=0.44\columnwidth,angle=0]{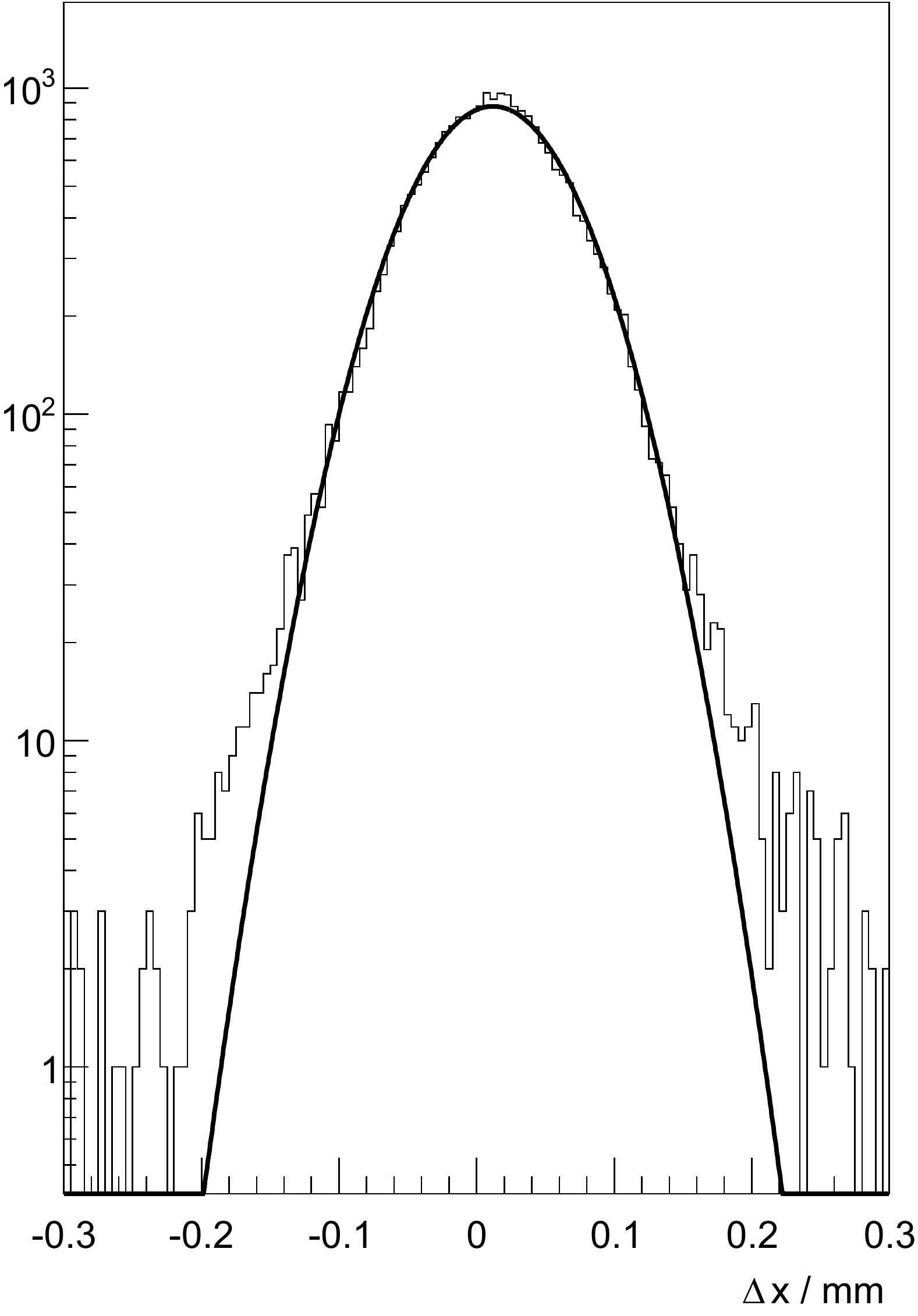}\\
\end{tabular}
\end{center}
\caption{Track residuals for the same array as in
fig.~\ref{fig:tb09_amps}, in normal (left) and logarithmic
(right) scale. The histogram has been fitted with a Gaussian.}
\label{fig:tb09_residuals}
\end{figure}

For the determination of the spatial resolution, in the direction
perpendicular to the fiber axes, the residuals
$\Delta{}x\equiv{}x_\mathrm{tr}-x_\mathrm{cl}$ are calculated. $x_\mathrm{tr}$ is the
intersection point of a particle as calculated using the track
information -- excluding, of course, the fiber layer under study -- 
and $x_\mathrm{cl}$ is the position
reconstructed in the fiber module and calculated according to
eq.~(\ref{eq:clusterwmean09}). Figure~\ref{fig:tb09_residuals} shows
an example of the distribution of residuals, for the same array as
above. A simple Gaussian fit is performed and the resulting standard
deviation is called the residual width.

\begin{figure}[htb]
\begin{center}
\includegraphics[width=\columnwidth,angle=0]{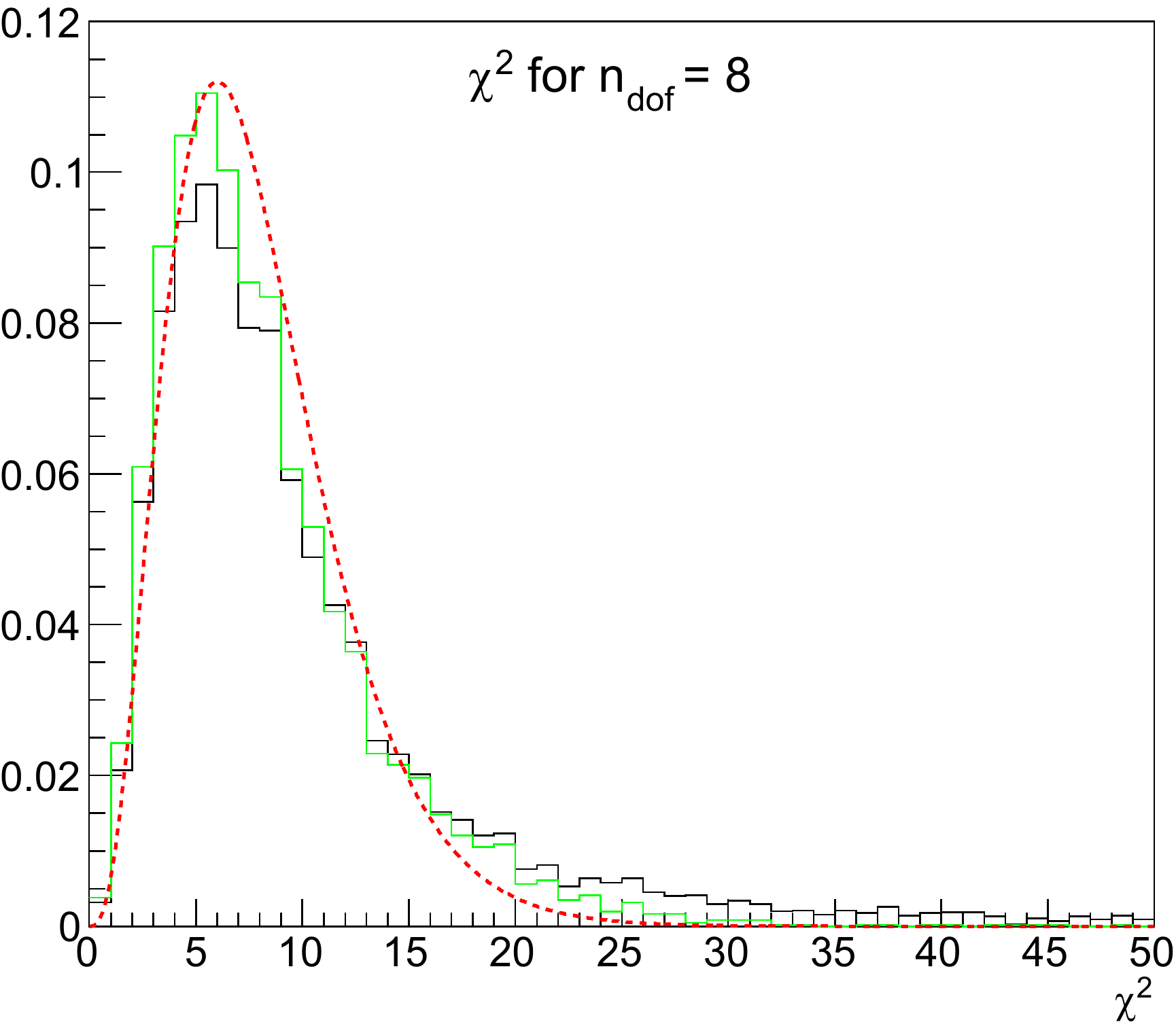}
\end{center}
\caption{Normalized $\chi^2$ distribution for tracks with $n_\mathrm{dof}=8$. The
black (dark) histogram is the distribution for tracks passing all quality
cuts described in the text. The green (bright) histogram is obtained if an
outlier rejection is performed in addition. For comparison, the red (dashed) curve is
the theoretical expectation ($\chi^2$ distribution for eight degrees
of freedom).}
\label{fig:tb09_chi2}
\end{figure}

Using the residual widths found in this way, the track fits can be
redone and the $\chi^2$ distribution for the track fit can be
calculated as a crosscheck (fig.~\ref{fig:tb09_chi2}).
Nice agreement with the theoretical expectation is found, especially
if an additional outlier rejection, removing clusters with a pull of
more than $4\sigma$, is performed. Here, the residual width of the
beam telescope modules in the $xz$-plane was assumed to be
$\unit[0.025]{mm}$. We checked that the results obtained for the spatial
resolutions of the fiber modules did not depend significantly on the beam telescope
resolution used in the track fit.

In order to obtain the underlying
spatial resolution of the fiber-SiPM compound from the residual width
shown above, the latter has to be
corrected for two effects. First, multiple Coulomb scattering leads to
a deviation of the particle trajectory from a straight line. Second,
due to the limited resolution of the silicon beam telescope and the
other fiber layers, the true track position is not perfectly
known. These effects have been included in a Monte Carlo study -- based
on the Geant4~\cite{ref:g4} package -- which models the material
budget present in the testbeam setup and the multiple scattering
process and uses the same tracking techniques as the testbeam
analysis, to calculate the necessary corrections.
\begin{figure}[htb]
\begin{center}
\includegraphics[width=\columnwidth,angle=0]{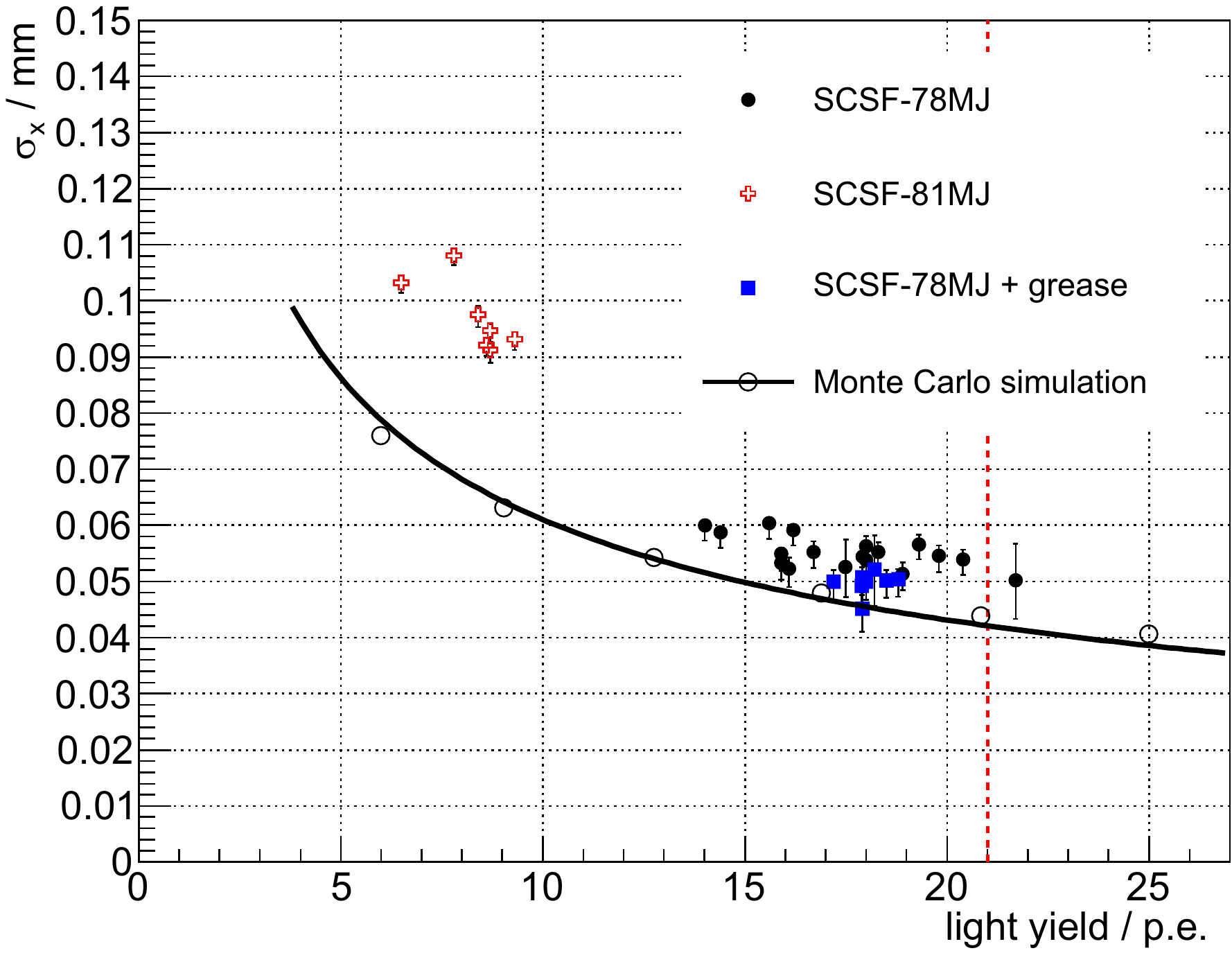}
\end{center}
\caption{Achieved spatial resolution of the tracker modules as a function of the light
yield. Each point corresponds to one SiPM array, using different markers for
different configurations of the tracker modules, regarding the fiber
type employed and the presence of
optical grease for the coupling of the scintillating fibers to the
SiPM.
The dashed line marks the expected light yield as
calculated in section~\ref{ref:discussion}. The prediction of the
Monte Carlo simulation described in section~\ref{ref:discussion} is
included, along with a fit to the results of the Monte Carlo
simulation according to eq.~(\ref{eq:mcfit}). The quoted light yields
are the median values of the respective amplitude
distributions. Taking the mean values instead would result in values
that are higher by $1\div2$ photo-electrons.}
\label{fig:tb09_finalPlot}
\end{figure}

Figure~\ref{fig:tb09_finalPlot} contains the main result of the
testbeam. Here, the light yield and spatial resolution that were
achieved for the prototype modules are shown, with each data point
corresponding to one silicon photomultiplier array. Different markers
are used to distinguish between arrays reading out modules made from
Kuraray SCSF-81MJ and SCSF-78MJ fibers, respectively, and to identify
those arrays with additional optical grease in front of them. The
correction for limited track resolution and multiple scattering has
been applied according to
\begin{equation}
\sigma_x^2=\sigma_\mathrm{res}^2-\sigma_\mathrm{MC}^2
\end{equation}
where $\sigma_\mathrm{res}$ is the residual width as shown for example in
fig.~\ref{fig:tb09_residuals}, $\sigma_\mathrm{MC}$ is the root mean
square deviation of the true track position from the reconstructed one
as found in the Monte Carlo simulation described above, and
$\sigma_x$ is the underlying spatial resolution of the
tracker module, plotted in fig.~\ref{fig:tb09_finalPlot}. We
conservatively assumed an intrinsic resolution of
$\unit[0.02]{mm}$ for the beam telescope ladders. Larger values, as
indicated by the $\chi^2$ distribution (fig.~\ref{fig:tb09_chi2}),
would increase the necessary correction and therefore result in lower
values of $\sigma_x$. A typical value is
$\sigma_\mathrm{MC}\cong18\,\upmu\mathrm{m}$. The error bars
shown in the figure are calculated as a quadratic sum of the
statistical and systematic uncertainties. The statistical uncertainties are
calculated from the $\chi^2$ variation in the Gaussian fits of the
residual distributions and are typically
$\lesssim\unit[1]{\upmu{}m}$, except for a few arrays that were located at the edge of the beam-spot,
which leads to a lower number of contained events and thus larger statistical uncertainties.
The systematical uncertainties are
estimated by varying the resolution of the beam telescope ladders
in the range from $\unit[0.01]{mm}$ to $\unit[0.03]{mm}$ and are
$\lesssim\unit[4]{\upmu{}m}$ for the arrays with the best spatial
resolution.

The results clearly show that a spatial resolution of better than
$\unit[0.05]{mm}$ has been achieved for some of the prototype
modules made from SCSF-78MJ fibers. At the same time, the light yield
is in the range from 14 to 20 photo-electrons (corresponding to mean
values of 15 to 22 photo-electrons) for most of the arrays. Despite the limited
statistics available, the figure also suggests that the optical grease
leads to an improved performance in terms of spatial resolution and a
more homogeneous light yield. Finally, there is a clear correlation
between light yield and spatial resolution, and the choice of fiber
material can affect the light yield by more than a factor of two. The
degradation of spatial resolution with decreasing light yield can be
understood from the statistical fluctuations inherent in the
calculation of the cluster position according to
eq.~(\ref{eq:clusterwmean09}).

For the high-yield fiber modules, i.e.~those containing SCSF-78MJ
fibers, the corresponding tracking efficiency varies between
$98.5\,\%$ and $99.5\,\%$. For the modules built with SCSF-81MJ
fibers, this value is worse, on the order of $90\,\%$.

\subsection{Discussion}
\label{ref:discussion}
To compare the results obtained in the previous section to
expectations, first the expected average light yield for a fiber ribbon can be
estimated as follows. On average, a minimum ionizing particle (MIP) deposits
$\unit[200]{keV/mm}$  in the scintillator material. From this energy
deposit, $8.25$ photons/MeV are generated. For the tracker geometry
presented here, five layers of $\unit[0.25]{mm}$ fibers with glue gaps
of $\unit[0.02]{mm}$, the average path length of a particle for perpendicular
incidence is calculated to be $\unit[0.69]{mm}$. Therefore, a MIP
creates 980 photons per fiber stack. Here and in the following,
we quote the median values of the distributions in order to be less
sensitive to tails in the distributions. The light trapping
efficiency of the fibers is defined by the refractive indexes of the
core and cladding materials and is $\unit[5.4]{\%}$ in each direction
along the fiber axes. In total, 105 photons are trapped inside the
fibers. Due to attenuation and reflection losses in the fibers and at
the mirrors at the far side of the fibers, respectively, only
$\unit[68]{\%}$ or 72 photons reach the readout SiPM. The fiber stack
measures $\unit[1.2]{mm}$ in height while the height of the SiPM array
is only $\unit[1.1]{mm}$. In addition, the Hamamatsu SiPM arrays are
protected by a glue layer of $\unit[0.275]{mm}$ thickness so that
photons emitted by a fiber near the fringe of the SiPM array may be
lost due to the angular emittance distribution. These two geometrical
effects lead to losses of $\unit[26]{\%}$, i.e.~only 53 photons reach
the sensitive area of the SiPM. The photon detection efficiency at an
overvoltage of $\unit[2]{V}$ is
$\unit[40]{\%}$ so that the expected average signal is $21$
photo-electrons. This value has been marked by the dashed line in
fig.~\ref{fig:tb09_finalPlot}.

We developed a Monte Carlo simulation of the detector to understand
the dependence of the spatial resolution on light yield. The
simulation of a passing particle starts with the generation of the
primary energy deposition, based on the path length in each fiber and
according to the appropriate distribution
that was extracted from Geant4~\cite{ref:g4}. The light-loss
effects detailed in the previous section are taken into account to
derive a mean number of photons for each fiber from which the actual
number of photons to be distributed onto the SiPM array is sampled
from a Poisson distribution. The response of the SiPM array is modeled
taking the Geiger-mode readout, electronic noise, as well as the effects of pixel
crosstalk and strip crosstalk into account. We use the appropriate
probabilities as determined in laboratory measurements
(fig.~\ref{fig:allNoise}) for an overvoltage of $\unit[2]{V}$, at which the SiPM arrays
were operated during the testbeam. In addition to these effects, we
expect a contribution to the detector resolution from optical
crosstalk between the fibers. As a first approach, this effect is included here
as an additional optical contribution to the electrical strip crosstalk probability
of $3\,\%$. The value for the optical crosstalk of $5\,\%$ is a free parameter
in the Monte Carlo simulation and has been adjusted to reproduce the
cluster profile properties as measured in the testbeam.
\begin{figure}[htb]
\begin{center}
\includegraphics[width=0.8\columnwidth,angle=0]{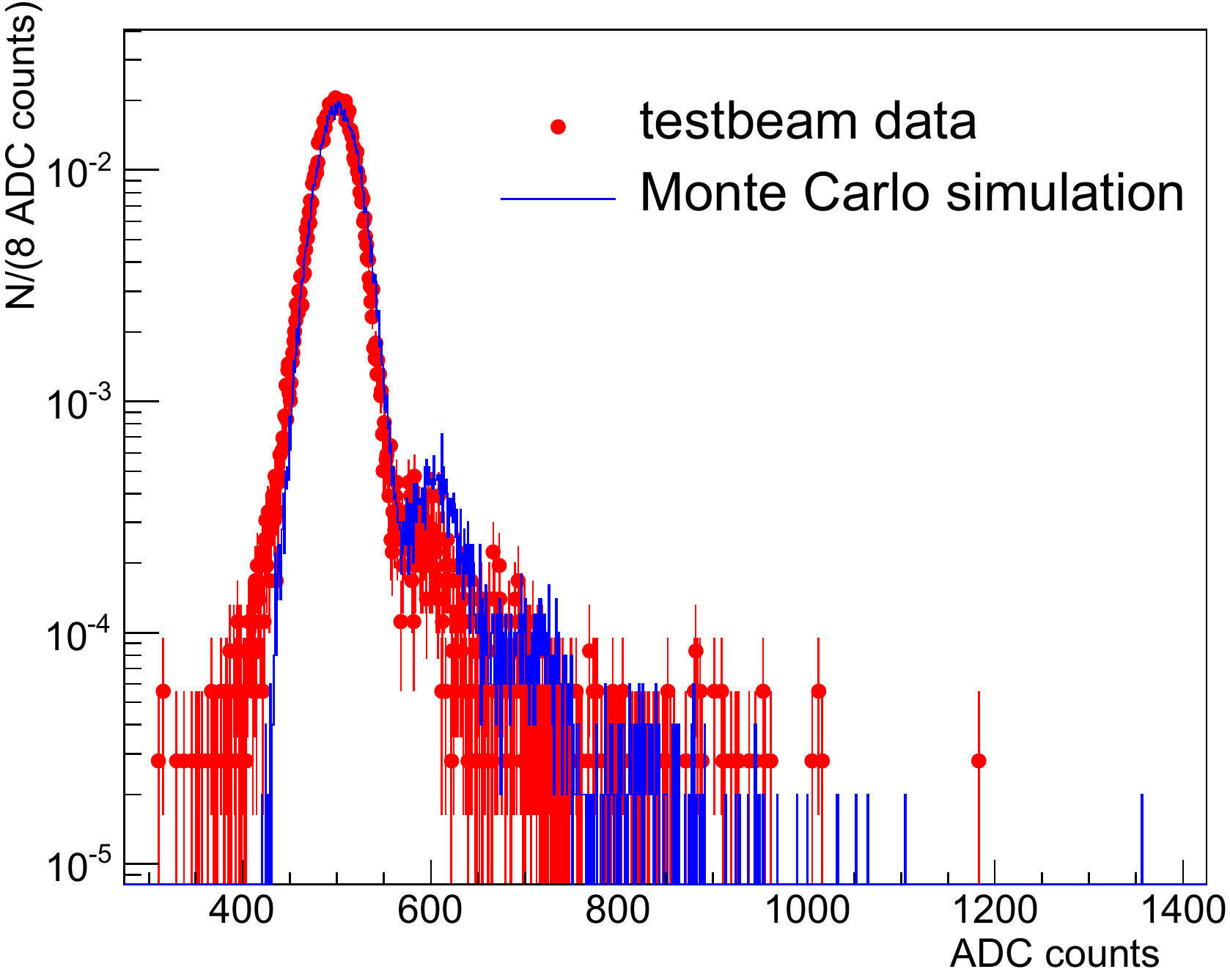}
\end{center}
\caption{Noise spectrum as measured for a typical Hamamatsu SiPM array
during the testbeam compared to the Monte Carlo simulation.}
\label{fig:noise_sim}
\end{figure}
\begin{figure}[htb]
\begin{center}
\includegraphics[width=0.8\columnwidth,angle=0]{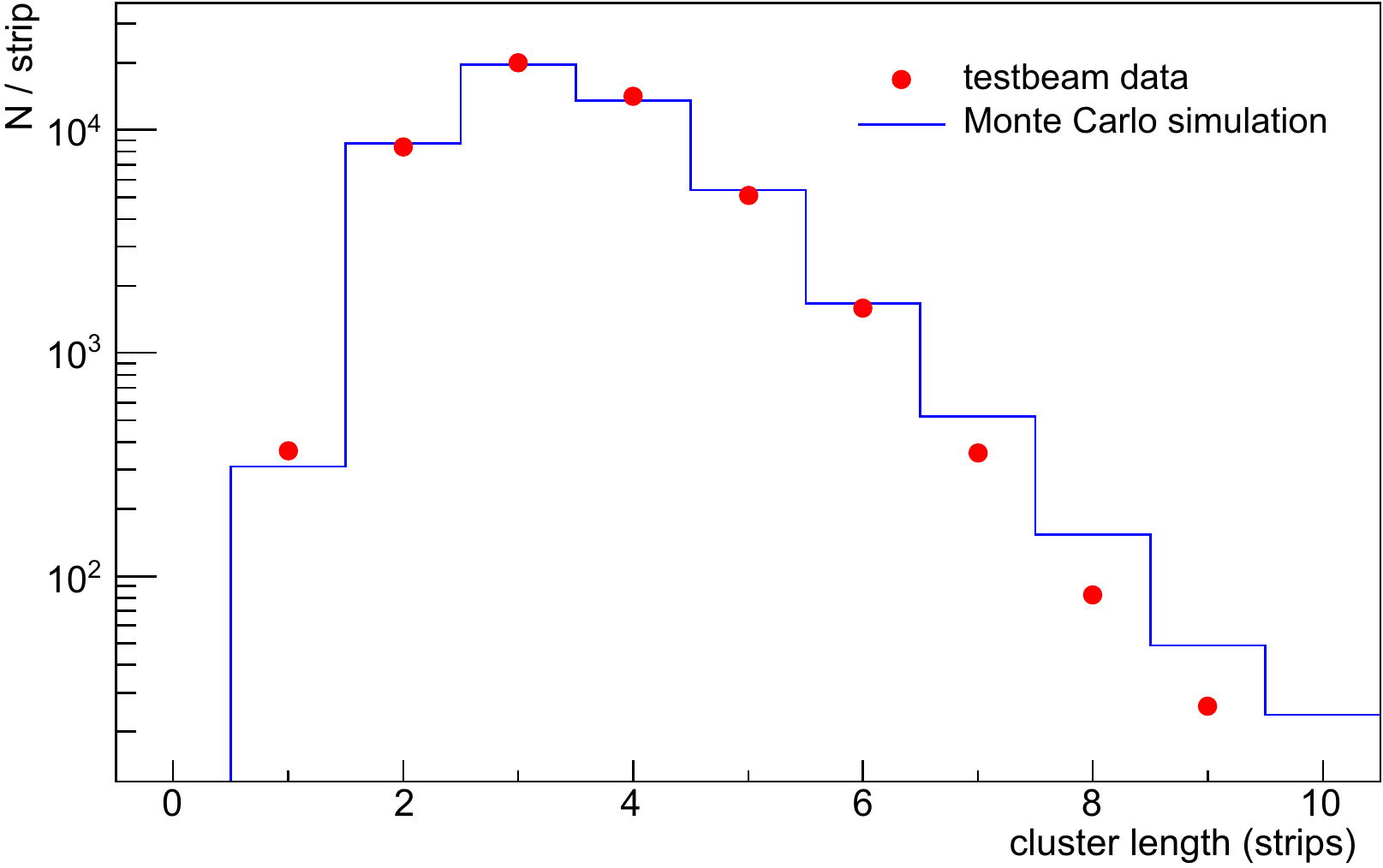}
\end{center}
\caption{Distribution of cluster length (number of SiPMs in array above threshold)
as obtained from testbeam data,
compared to the Monte Carlo simulation.}
\label{fig:length_sim}
\end{figure}
\begin{figure}[htb]
\begin{center}
\includegraphics[width=0.8\columnwidth,angle=0]{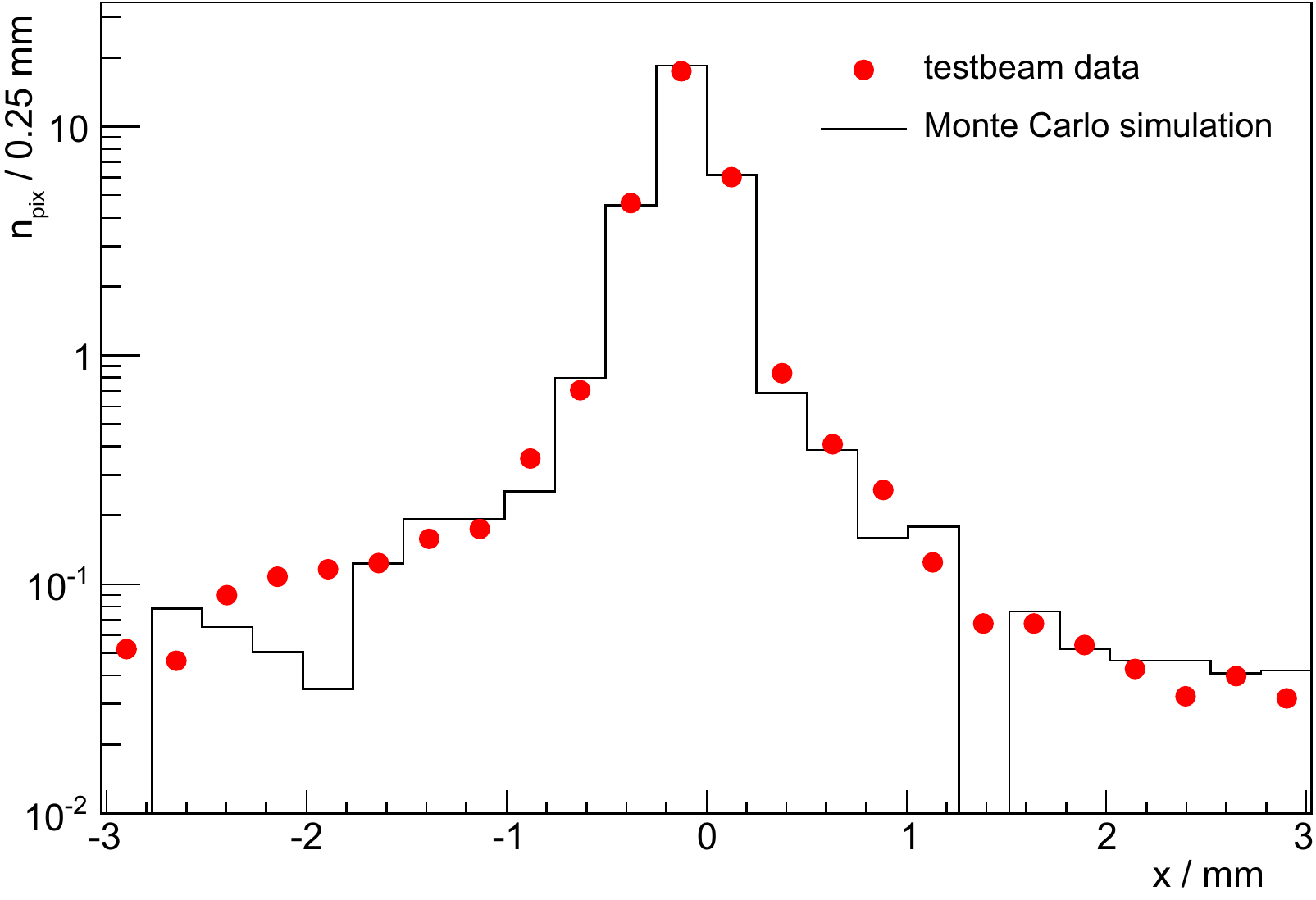}
\end{center}
\caption{Average signal cluster as obtained from testbeam data,
compared to the Monte Carlo simulation.}
\label{fig:cluster_sim}
\end{figure}
Finally, fake hits are added according to the fake rate measured in the
laboratory (fig.~\ref{fig:allNoise}). To illustrate the overall good
description, we compare a typical noise spectrum, the distribution of cluster length,
and the average
signal cluster for testbeam data and simulation in figures
\ref{fig:noise_sim}, \ref{fig:length_sim}, and \ref{fig:cluster_sim}, respectively. The
latter represents an especially important crosscheck, as the cluster
shape depends crucially on all the effects mentioned above and at the
same time, it is the basis for the calculation of the particle
intersection point according to eq.~(\ref{eq:clusterwmean09}).

The prediction of the Monte Carlo simulation for the detector
resolution as a function of light yield is included in
figure~\ref{fig:tb09_finalPlot}. For the light yield range relevant to
our study, the results for the
detector resolution $\sigma_x$ in the simulation can be described
as
\begin{equation}
\label{eq:mcfit}
\sigma_x=a/\sqrt{n_\mathrm{pe}}
\end{equation}
where $n_\mathrm{pe}$ is the light yield.
A fit to the simulation results for SCSF-78MJ fiber modules
with optical grease yields a value of
$a=\unit[0.193]{mm}$. It can be seen that the prediction of the Monte
Carlo simulation is too optimistic by about $\unit[0.005]{mm}$ for the
high-yield modules, and the resolution of the low-yield module is
underestimated by roughly $\unit[20]{\%}$, but the trend of the data
is well reproduced.

\section{Outlook}
\label{sec:outlook}
An important issue that limits the spatial resolution achievable in
the current configuration is the fact that
the distribution of photons on the SiPM array is smeared out over several
channels because of the optical gap between SiPM array and fiber. The
photons have to bridge a distance
of $\unit[275]{\upmu{}m}$ defined by the optical glue that Hamamatsu
puts on their MPPC 5883 SiPM devices, to protect them against
environmental and handling damage. While the
influence of crosstalk and the closely linked strip crosstalk as well
as the total detected light yield of the fibers play a dominant role
for the achieved performance of the detector module, the thickness of
the glue layer on top of the SiPM devices can be modified much more easily
since it does not require a complete redesign of the SiPM.

Devices with a reduced glue layer of $\unit[0.1]{mm}$ have already been delivered
by Hamamatsu and should allow for a position resolution of $\unit[0.04]{mm}$. 
A new generation of 64-channel SiPM arrays with no additional glue layer has been designed 
and will become available in summer 2010. In addition, the optical coupling 
has been optimized further. Both effects together
should allow for a position resolution of $\unit[0.035]{mm}$ according to
Monte Carlo simulations. It is also obvious that optimized software
algorithms which take the discrete structure of the detector into
account could lead to further significant improvements. Finally,
Monte Carlo simulations show that the optimal readout pitch for
a fiber detector constructed from $\unit[0.25]{mm}$ fibers would be $\unit[0.125]{mm}$. 
In addition, the new 64-channel SiPM arrays would also allow a double-sided
readout of the fibers. These measures together would allow the
construction of scintillating fiber tracker modules with high
intrinsic redundancy and an ultimate position resolution of better than $\unit[0.02]{mm}$
according to Monte Carlo simulations.

\section*{Acknowledgments}
We gratefully acknowledge the generous support from R.~Battiston,
G.~Ambrosi and P.~Azzarello of the INFN Perugia who allowed
us to use tracker ladders from the AMS-02 project for the beam
telescope. We thank CERN for valuable beam time and support during the testbeam measurements.
The measurements of the angular emission spectra were
performed at the Georg Simon Ohm college in Nuremberg. The setup for the determination of the fill
factor is based on an idea pioneered by H.G.~Moser at the Max-Planck-Institut f\"ur Physik in Munich.

\end{document}